\documentclass{nature}

\usepackage{graphicx}
\usepackage{epstopdf}
\usepackage{amsmath}
\usepackage{epsfig}
\usepackage{subfig}
\usepackage{float}
\usepackage{multirow}
\usepackage{amssymb}
\usepackage{tikz}
\usepackage{braket}
\usepackage{stackengine}
\usepackage[export]{adjustbox}
\usepackage{xr}
\newcommand\LiHoF{LiY$_{1-x}$Ho$_x$F$_4$}
\newcommand{\onlinecite}[1]{\hspace{-1 ex} \nocite{#1}\citenum{#1}}

\makeatletter
\let\saved@includegraphics\includegraphics
\AtBeginDocument{\let\includegraphics\saved@includegraphics}
\renewenvironment*{figure}{\@float{figure}}{\end@float}
\renewenvironment*{table}{\@float{table}}{\end@float}
\makeatother

\title{Long-lived divergence from equilibrium of electrons, nuclear spins and lattice for a solid state ion trap at low temperature}

\author{Guy Matmon$^1$, Manuel Grimm$^{1,2}$, Markus M\"uller$^1$, Byron\,J. Villis$^{3,4}$, Andrew\,J. Fisher$^{3,5}$ and Gabriel Aeppli$^{1,2,6}$
}

\begin{document}

\maketitle

\begin{affiliations}
	\item Paul Scherrer Institute, Forschungstrasse 111, 5232 Villigen, Switzerland
	\item Department of Physics and Quantum Center, ETH Z\"urich, 8093 Z\"urich, Switzerland
	\item London Centre for Nanotechnology, University College London, 17--19 Gordon St., London, WC1H 0AH, United Kingdom
    \item Photon Foundry Pty Ltd, NSW 2130, Australia
	\item Department of Physics and Astronomy, University College London, Gower Street, London, WC1E 6BT, United Kingdom
	\item Institute of Physics, EPF Lausanne, 1015 Lausanne, Switzerland
\end{affiliations}

\begin{abstract}
A fundamental problem in physics as well as engineering is equilibration. For example, the regulation of thermal and quantum fluctuations enables thermal and quantum annealing of complex systems. A key question is how different subsystems, such as electrons, nuclear spins and phonons equilibrate on their own as well as with each other. Level crossings play a special role in the dynamics of coupled degrees of freedom, for it is here that entanglement can be maximized to speed up relaxation. Here we use optical methods to establish the electronuclear level scheme, including avoided and unavoided crossings, and to examine equilibration of a rare earth ion (Ho$^{3+}$) in a salt (LiYF$_4$), a model system with quantum fluctuations which can be tuned via an external magnetic field transverse to the crystallographic long axis of the tetragonal host~\cite{Bitko:1996zr,Wu:1991,Brooke:1999fk}. We track the state of the system by monitoring the populations of the levels as a function of swept longitudinal fields, and discover that at low temperatures, depending on the experimental protocol, vastly different non-equilibrium states arise. We find evidence that nuclear spin excitations diffuse and equilibrate without the assistance of phonons, and tend to acquire higher effective temperatures than the electronic spins and the yet cooler lattice. A theory of thermally assisted tunneling rules out the standard scenario of phonon -assisted tunneling, and instead  suggests that fast dynamics at level anti-crossings are facilitated entirely by nuclear spin diffusion. This provides a new understanding of thermalization and related slow relaxation phenomena in dense, multi-component interacting quantum systems\cite{Malkin:2005qf,Barbara:1999fk}.

\end{abstract}

\noindent Ions in salts are the solid state analogs of ions in traps. They also realize in a tunable way many of the models of interacting two level systems at the heart of quantum many-body physics, whose purview ranges from quantum phase transitions in the clean limit\cite{Bitko:1996zr} and (quasi-)many-body localization in disordered interacting systems\cite{Basko:2006uq, Nandkishore:2015vr, Beckert:2024pi}, to quantum entanglement among primary and spectator degrees of freedom\cite{Ronnow:2005}. Different ions access different electronic and nuclear spin levels, both including exceptionally long-lived excited states\cite{Thiel:2011}. The magnetic dipolar interactions between the ions can be regulated by their density\cite{Youngblood:1982} as well as via optical excitation to higher lying electronic states, leading to proposals for gate-based quantum computing\cite{Grimm:2021db,Kinos:2021}. 

\noindent Many experiments are well understood because they are either in the limit of equilibrium quantum statistical physics where we can apply the concept of a single temperature characterizing the lattice, electron and nuclear spin fluctuations, or in the limit of atomic physics where the electronuclear degrees of freedom are coupled to the outside world only on account of electromagnetic fields manipulated or observed remotely. Nonetheless, there are also experiments which remain puzzling because they seem to fall into neither category.  For example, for one system, LiY$_{1-x}$Ho$_x$F$_4$, there is an apparent crossover between "antiglass", where quantum fluctuations increasingly dominate on cooling and where extraordinarily narrow holes can be burned into the low energy magnetic spectrum, and classical spin glass states as the contact of the lattice to an external bath increases\cite{Silevitch:2010eu}. This result suggests that the lattice and electronuclear temperatures might differ over extended windows of time and allow certain excitations to survive for even longer, notwithstanding the relatively high density of ions. In order to scrutinize this possibility we set out to map the electronuclear states in LiY$_{1-x}$Ho$_x$F$_4$ as a function of external bath temperature, longitudinal field history and quantum mixing, which can be easily regulated using an external transverse field. We have acquired the relevant optical data in the challenging sub-Kelvin regime where single ion quantum magnetism has been probed previously only by SQUID magnetometry, which is essentially blind to the nuclear spins\cite{Giraud:2001vn} and their extremely slow dynamics. Here instead we monitor directly how they are driven out of equilibrium. The level crossings implicated in the electronuclear relaxation are measured directly as a function of transverse field $B_x$ and categorized as linear or quadratic in $B_x$. Examination of the optical amplitudes and polarization rotation reveals that a slow sweep through the hysteresis loop establishes three distinct effective temperatures, in a hierarchy where the lattice is cold, the electrons luke-warm and the nuclear spins hot. 

\subsection{Symmetry and energy levels} Ho$^{3+}$ is a non-Kramers ion with a $4f$ shell containing 10 electrons, and Fig. \ref{Fig:Levels_Trans_Field} illustrates its energy levels when substituted for non-magnetic  Y$^{3+}$ in LiYF$_4$. Strong spin-orbit coupling separates electronic  states of different total angular momentum $J$ by hundreds of meV. Each $J$-manifold is further split by the electric potential of the host crystal to form a set of crystal-field (CF) states, typically tens of meV apart. Fig. \ref{Fig:Levels_Trans_Field} shows the lowest CF states of the two lowest spin-orbit manifolds. Due to the $S_4$ symmetry at the Ho sites, the electronic CF states are non-magnetic singlets or magnetic doublets. The Ho ground state is a doublet which carries a large Ising moment along the \mbox{$c$-axis} ($\mu \approx 6.9\,\mu_\mathrm{B}$), and we label these two degenerate states as $\ket{\uparrow}, \ket{\downarrow}$. The strong  hyperfine (HF) interaction orients the Ho nuclear spin ($I=7/2$) along the electronic moment, splitting the electronic doublet into eight nearly equidistant (separated by $E_\mathrm{HF} \approx 18\,\mu$eV, $0.145\,\text{cm}^{-1}$) doubly-degenerate HF states. The Hamiltonian of a single ion is discussed in more detail in the Methods' section~\ref{Sec:Spectrum}. The degeneracy of these HF doublets can be lifted by breaking the time-reversal symmetry with a longitudinal magnetic field, see Figs.~\ref{Fig:Long_Field_0_Trans} and \ref{Fig:Long_Field_5000_Trans}. We probe these 16 HF states via optical transitions to higher-lying singlet CF states indicated in Fig.~\ref{Fig:Levels_Trans_Field}, using infrared (IR) optical absorption as \textit{e.g.} in Refs.~[\,\onlinecite{Matmon:2016fk,Boldyrev2019}]. Note that the highest transition energy corresponds to the excitation from the lowest energy doublet state.

\subsection{Simultaneous low-temperature spectroscopy and magnetometry}
\begin{figure}
	\centering
	\begin{tabular}{ccc}
        \hspace{-24mm}
		\subfloat{
			\stackinset{l}{-5pt}{t}{0pt}{(a)}{\includegraphics[width =0.23\linewidth,valign=m]{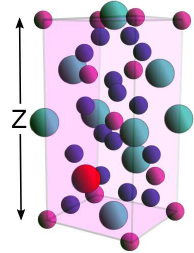}}
			\label{Fig:Sample}
		}
        &
		\hspace{-48mm}
		\subfloat{
			\stackinset{l}{-7pt}{t}{0pt}{(b)}{\includegraphics[width =0.26\linewidth,valign=m]{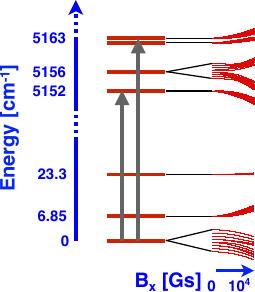}}
			\label{Fig:Levels_Trans_Field}
		}&
		\hspace{-27mm}
		\subfloat{
			\stackinset{l}{0pt}{t}{-18pt}{(c)}{\includegraphics[width =0.45\linewidth,valign=m]{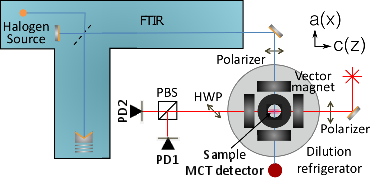}}
			\label{Fig:Exp_Setup}
		}\\
        \hspace{-7mm}
		\subfloat{
			\stackinset{l}{37pt}{t}{5pt}{(d)}{\includegraphics[width =0.39\linewidth,valign=b]{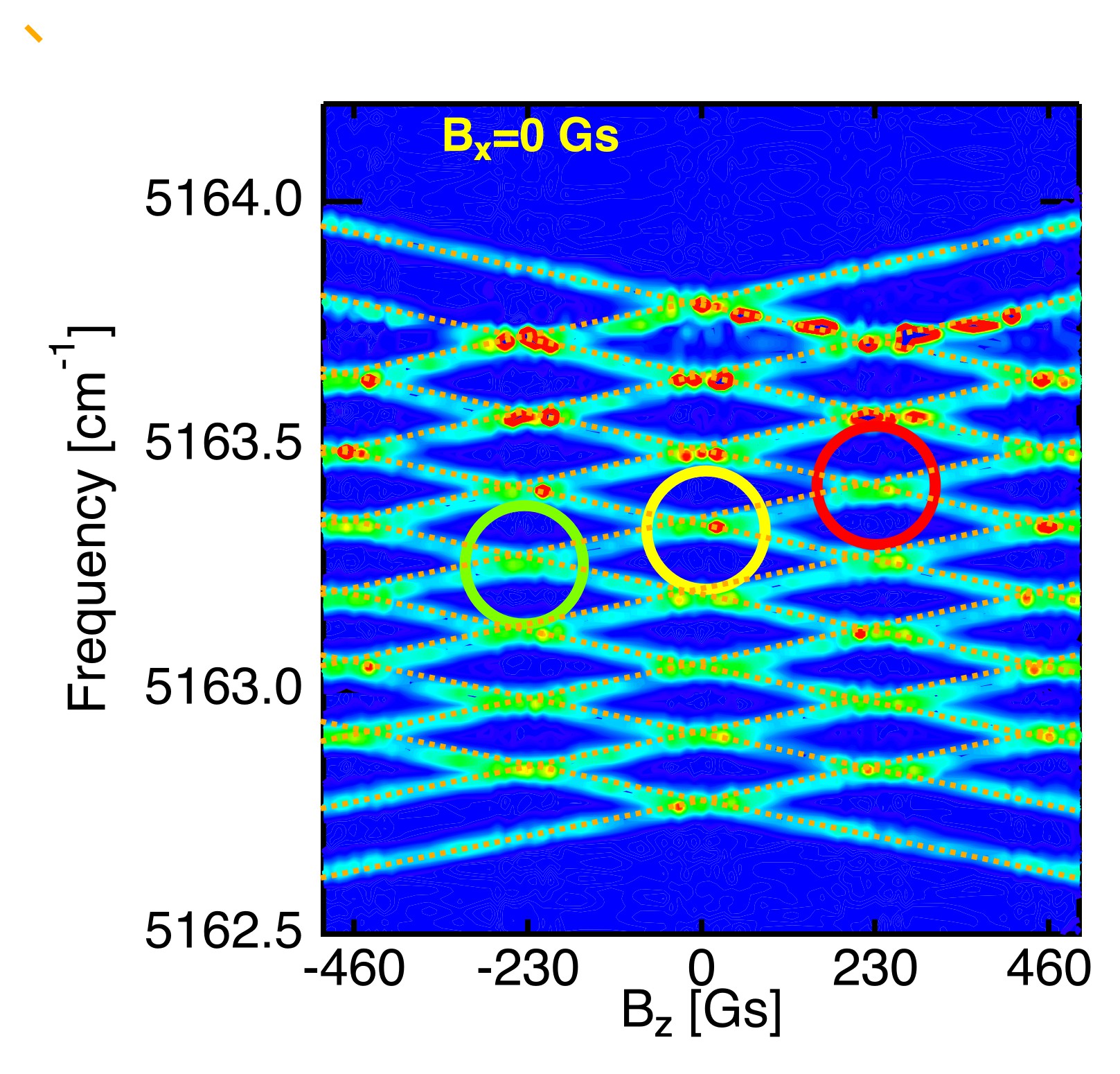}}
			\label{Fig:Long_Field_0_Trans}
		}&
        \hspace{-11mm}
		\subfloat{
			\stackinset{l}{-8pt}{t}{-5pt}{(e)}{\includegraphics[width =0.295\linewidth,valign=b]{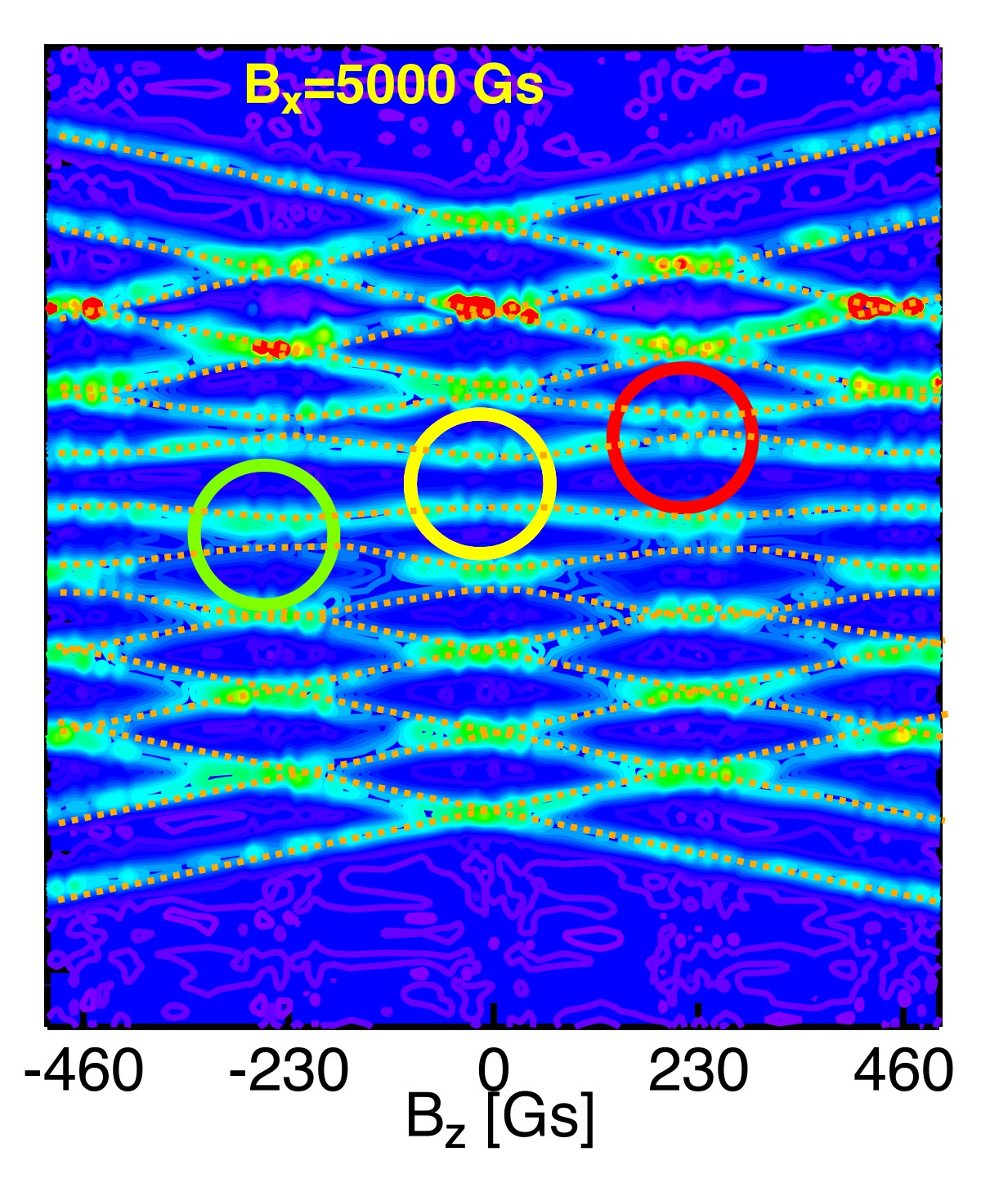}}
			\label{Fig:Long_Field_5000_Trans}
		}&
        \hspace{-11mm}
		\subfloat{
			\stackinset{l}{-8pt}{t}{-8pt}{(f)}{\includegraphics[width =0.33\linewidth,valign=b]{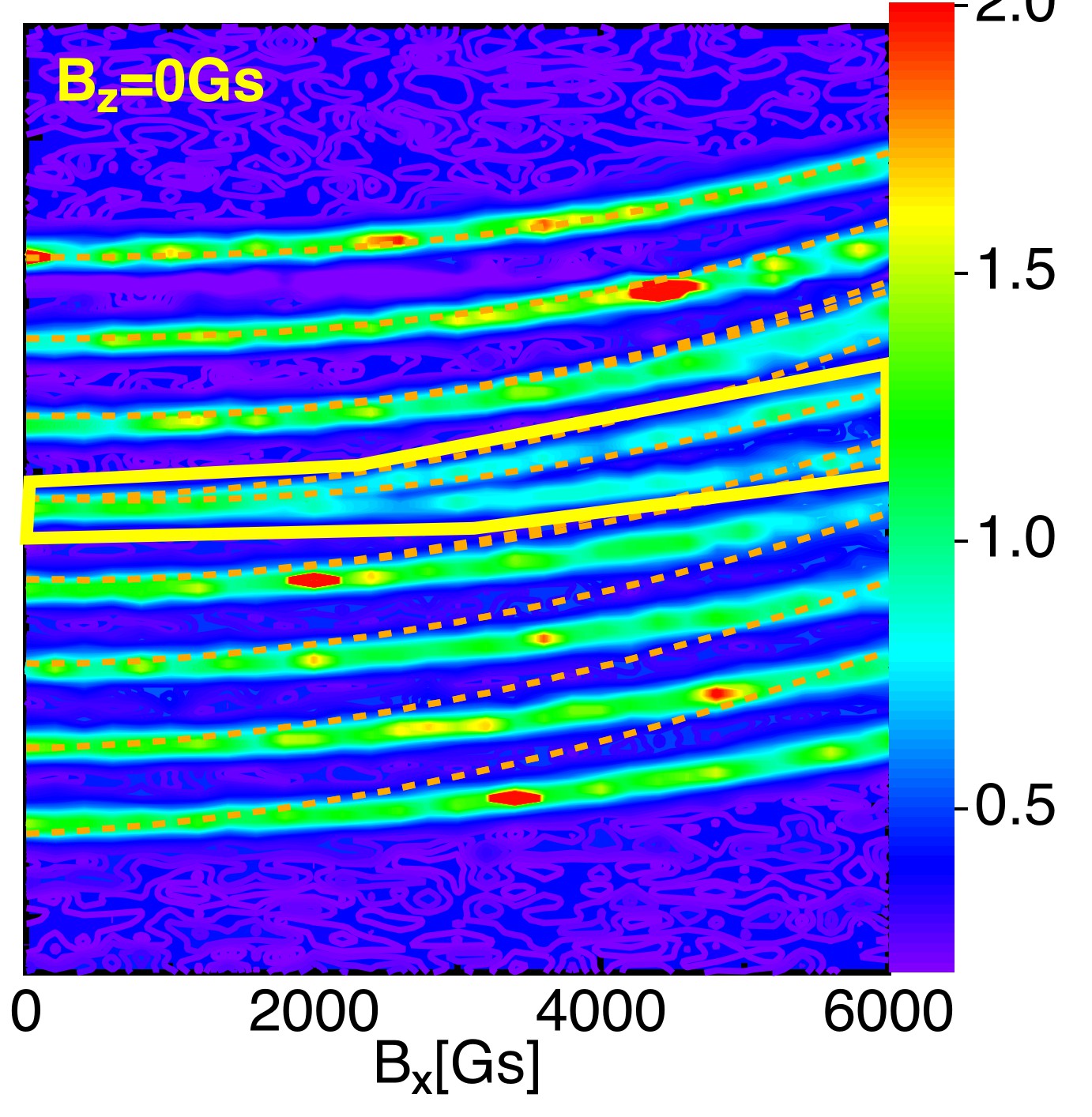}}
			\label{Fig:Trans_Field_0_Long}
		}
	\end{tabular}
	\caption{Energy levels and setup. \protect\subref{Fig:Sample} A LiY$_{1-x}$Ho$_x$F$_4$ unit cell with a Ho$^{3+}$ ion in red. \protect\subref{Fig:Levels_Trans_Field} The lowest CF levels in the two lowest spin-orbit manifolds ($^5I_8$ and $^5I_7$) in a transverse field. The grey arrows show the transitions used in this work. \protect\subref{Fig:Exp_Setup} The experimental setup. \protect\subref{Fig:Long_Field_0_Trans} and \protect\subref{Fig:Long_Field_5000_Trans} HF levels in a longitudinal magnetic field sweep with different transverse fields, at 10\,K. \protect\subref{Fig:Trans_Field_0_Long} Measured HF levels in a transverse magnetic field sweep. The color axis units are absorbance. The dotted orange lines are the calculated HF levels. The circles in \protect\subref{Fig:Long_Field_0_Trans} and \protect\subref{Fig:Long_Field_5000_Trans} and the polygon outlined in yellow in \protect\subref{Fig:Trans_Field_0_Long} demonstrate the gap openings calculated in Fig. \ref{Fig:Trans_Field_Gaps}.}
	\label{Fig:Introduction}
\end{figure}

A dilute \LiHoF\ crystal ($x=0.003$) was mounted in a dilution refrigerator (base temperature 20\,mK) with optical access along the crystal $a$ and $c$ axes, and a vector magnet. We monitored the HF levels and their occupation by recording the infrared absorption spectrum along the $a$-axis ($E\parallel c$) with a Fourier-transform infrared (FTIR) spectrometer, while we measured  Faraday rotation, which measures only the electronic moments, with a visible laser beam along the $c$-axis to establish the magnetization. Fig. \ref{Fig:Exp_Setup} and a description in Methods Section \ref{sec:Experimental} provide details on the experimental setup.	

\begin{figure}
	\centering
	\begin{tabular}{ccc}
        \hspace{0mm}
		\subfloat{
			\stackinset{l}{-8pt}{t}{-10pt}{(a)}{\includegraphics[width =0.35\linewidth,valign=b]{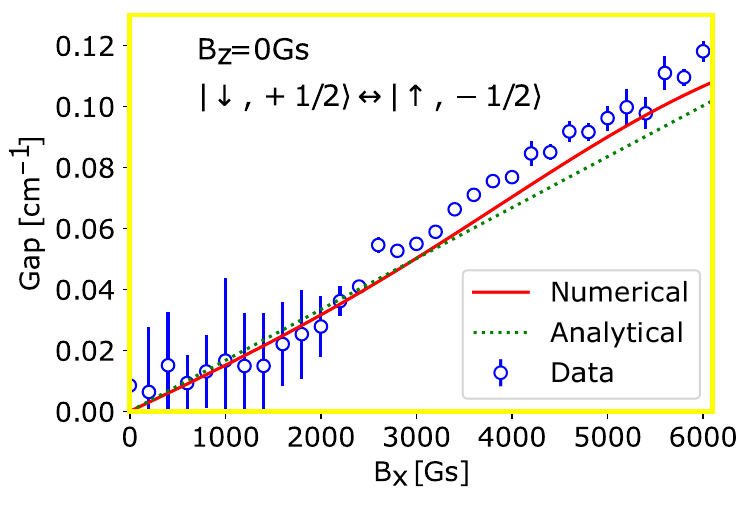}}
			\label{Fig:Gap_0_78}
        }&
        \hspace{-7mm}
		\subfloat{
			\stackinset{l}{-8pt}{t}{-10pt}{(b)}{\includegraphics[width =0.30\linewidth,valign=b]{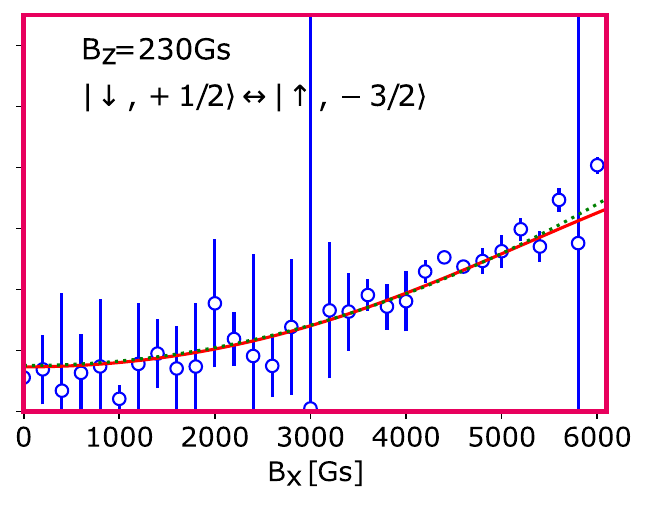}}
			\label{Fig:Gap_230_67}
        }&
        \hspace{-7mm}
		\subfloat{
			\stackinset{l}{-8pt}{t}{-10pt}{(c)}{\includegraphics[width =0.30\linewidth,valign=b]{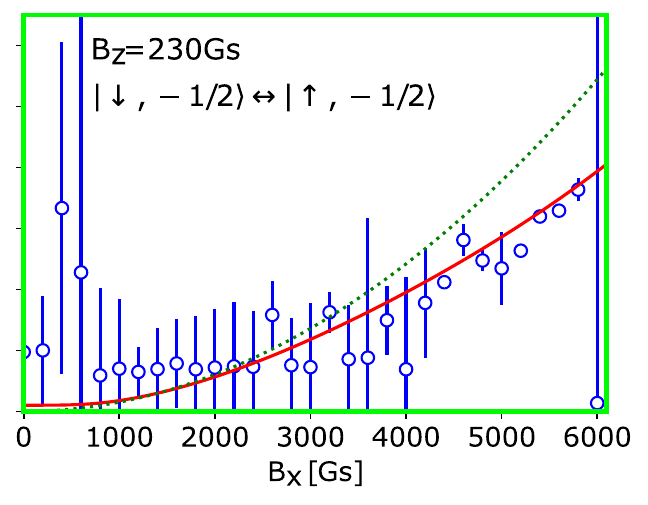}}
			\label{Fig:Gap_230_89}
            }
	\end{tabular}
	\caption{Calculated gaps in a transverse field. \protect\subref{Fig:Gap_0_78} The evolution of gaps between HF levels as a function of transverse field with $\lvert\Delta I_z\rvert=1$ (linear in $B_x$). The green dotted line is the analytical prediction, the red line is the leading order perturbation theory, see Methods Section~\ref{Sec:Spectrum}. The deviations are due to neglected higher order corrections in $B_x$. \protect\subref{Fig:Gap_230_67} $\lvert\Delta I_z\rvert=2$ at $B_z=230$\,Gs, (quadratic in $B_x$ + offset). \protect\subref{Fig:Gap_230_89} $|\Delta I_z| = 0$ (quadratic in $B_x$).}
	\label{Fig:Trans_Field_Gaps}
\end{figure}

\subsection{Ground-state level spectrum}	 
Fig.~\ref{Fig:Trans_Field_0_Long} shows the measured and calculated absorption lines as a function of transverse field. The small deviations between the numerical and experimental results are due to neglect of quadrupolar HF corrections to the doublet states, and of HF corrections to the final singlet state. As the final state in the $^5I_7$ manifold we use the singlet at energy $5163\,{\rm cm}^{-1}$, as it only acquires a minimal moment in response to a transverse field, and thus does not introduce a noticeable additional HF splitting. Additional measurements of the longitudinal and transverse field dependence appear in Figs.~\ref{Fig:Long_Field_Inset} -- \ref{Fig:Trans_Field_SI}. 

\noindent The longitudinal field component couples to the large Ising moments, which induces crossings of HF levels at fields that are multiples of \mbox{$\Delta B_z \approx 230$\,Gs}. In the absence of a transverse field (Fig. \ref{Fig:Long_Field_0_Trans}, see also Ref.~[\,\onlinecite{Boldyrev2019}]), most of these crossings are protected since the longitudinal field preserves the crystalline $S_4$ symmetry, preventing the hybridization of states with different $S_4$ quantum numbers. The only exceptions are the (anti-)crossings between HF states with a difference $\lvert\Delta I_z\rvert \mod 4 = 2 $ (\textit{i.e.},\ $\lvert\Delta I_z\rvert = 2,6$) in nuclear spin projection. Those share the same symmetry and hybridize because the transverse HF interaction couples them via the admixture of higher-lying CF states, see Sec.~\ref{Sec:Spectrum} in the Methods for details. The larger the $\lvert\Delta I_z\rvert$, the smaller the gaps are. 

\noindent A finite transverse field breaks both the time-reversal and the $S_4$ symmetry and thus lifts all degeneracies. We can classify the HF crossings by the scaling of the gap with a (small) transverse field. The $S_4$ symmetry dictates that level crossings with odd $\Delta I_z \mod 4 \in \{1,3\}$ open up at first order in $B_x$ (Fig. \ref{Fig:Gap_0_78}), whereas crossings with even \mbox{$\Delta I_z \mod 4 =0 $} come with gaps quadratic in $B_x$.

\subsection{Dynamics}

 One of the key features of LiHo$_x$Y$_{1-x}$F$_4$ is that quantum speed-up is easily regulated by modest external transverse fields\cite{Bitko:1996zr,Wu:1991,Brooke:1999fk}. Particularly important is the electronuclear entanglement indicated by the hitherto unmeasured gaps at the avoided level crossings: quantum tunneling between different electron spin states depends on the same matrix elements that determine these gaps. Avoided level crossings imply the mixing of electronuclear wavefunctions, as for example between $\ket{\downarrow,I_z=+1/2}$ and $\ket{\uparrow,I_z=-1/2}$ highlighted in Figs. \ref{Fig:Gap_0_78} and \ref{Fig:Trans_Field_0_Long}, which facilitates the adiabatic, resonant flipping of electron spins upon slowly approaching the crossings. High-resolution spectroscopic data of the type shown in Fig. \ref{Fig:Introduction} provide not just the eigenenergies near the level crossings, but, via analysis of peak intensities, estimates of all state populations, which can yield key physical parameters such as electronic and nuclear magnetization, and effective temperatures of the electronic and nuclear spins. If collected at appropriate speed, they can establish the occupancy evolution of all sixteen states in the electronuclear Hilbert space, which is particularly pronounced, for example, where there are level crossings with tunable gaps. However, data collection can be slow on the scale of interesting time-dependent effects, and also results in (IR) photo-induced sample heating. We adopted two strategies to deal with these shortcomings: the first by tracking the net electronic magnetization via off-resonant optical polarization rotation (Faraday effect), and the second by reducing the FTIR photon flux.

\begin{figure}[!ht]
	\centering
	\begin{tabular}{ccc}
	\hspace{-9mm}
    \subfloat{
		\stackinset{l}{68pt}{t}{3pt}{(a)}{\includegraphics[width = 0.345\linewidth,valign=t]{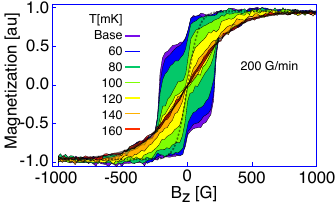}}
		\label{Fig:Hyst_Temp}
    }&
	\hspace{-6mm}
	\subfloat{
		\stackinset{l}{55pt}{t}{3pt}{(b)}{\includegraphics[width = 0.32\linewidth,valign=t]{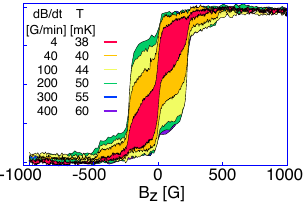}}
		\label{Fig:Hyst_Sweep}
	}&
	\hspace{-6mm}
	\subfloat{
		\stackinset{l}{60pt}{t}{3pt}{(c)}{\includegraphics[width = 0.32\linewidth,valign=t]{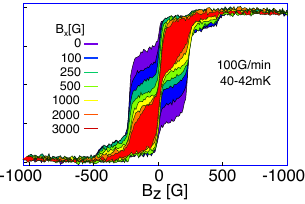}}
		\label{Fig:Hyst_Trans}
	}\\
	\hspace{-9mm}
    \subfloat{
		\stackinset{l}{68pt}{t}{6pt}{(d)}
        {\includegraphics[width = 0.345\linewidth,valign=t]{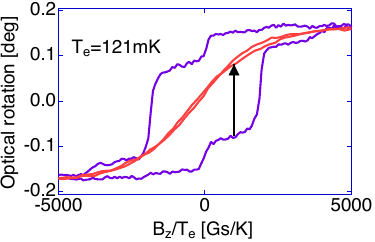}}
		\label{Fig:Spin_Relaxation_Hyst}
	}&
    \hspace{-13mm}
    \subfloat{
        \stackinset{l}{70pt}{t}{6pt}{(e)}{\includegraphics[width = 0.335\linewidth,valign=t]{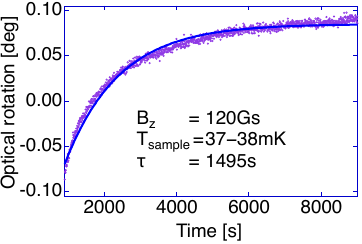}}
        \label{Fig:Spin_Relaxation}
    }&
	\hspace{-10mm}
    \subfloat{
        \stackinset{l}{75pt}{t}{6pt}{(f)}{\includegraphics[width = 0.35\linewidth,valign=t]{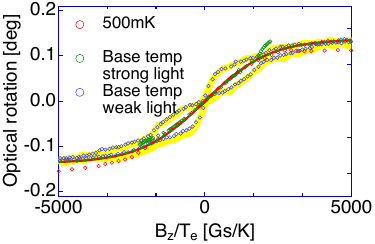}}
        \label{Fig:Faraday_Magnetization}
    }
	\end{tabular}
	\caption{Magnetization measurements. \protect\subref{Fig:Hyst_Temp} -- \protect\subref{Fig:Hyst_Trans} Hysteresis of the magnetization when sweeping the longitudinal field for different temperatures, sweep rates and transverse fields, respectively. Overlaid in grey in \protect\subref{Fig:Hyst_Temp} are paramagnetic equilibrium magnetization curves at 40\,mK (dashed) and 160\,mK (solid). \protect\subref{Fig:Spin_Relaxation_Hyst} A magnetization curve measured at base temperature, without light, with a transverse field of 0\,Gs and 10$^4$\,Gs (violet and red, respectively). The arrow at 120\,Gs shows the difference in magnetization in and out of equilibrium. \protect\subref{Fig:Spin_Relaxation} Relaxation of magnetization measured over time at 120\,Gs, without a transverse field. \protect\subref{Fig:Faraday_Magnetization} Faraday rotation vs. effective field ($B_z/T_e$) during FTIR measurements. Red is a high resolution and strong illumination sweep with the thermostat set to 500\,mK. Green is with strong illumination, without heating ($T_\text{sample}$ 102\,mK). Blue is low resolution with weak illumination ($T_\text{sample}\approx$50\,mK). The red and green lines are fits corresponding to $T_e$ of 507\,mK and 225\,mK, respectively. The yellow outline is the hysteresis curve measured with no light at 120\,mK, as shown in \protect\subref{Fig:Hyst_Temp}, which tracks the Faraday-measured (blue) magnetization data best.}
	\label{Fig:Hyst}
\end{figure}

\subsection{Faraday Rotation} Fig.~\ref{Fig:Hyst} shows the Faraday rotations, directly proportional to the electronic magnetization, for different temperatures, sweep rates, and transverse fields, without IR illumination. Above a temperature of \mbox{$T\approx160$\,mK}, the behaviour in Fig.~\ref{Fig:Hyst_Temp} is that of paramagnetic Ho$^{3+}$ ions in thermal equilibrium, with magnetization \mbox{$M \propto \tanh\left(\mu B_z/k_B T \right)$}. Below that temperature, the electronic spin relaxation slows down dramatically (presumably as comparatively rapid Raman relaxation dies off) and we observe an increasing deviation from thermal equilibrium with large step-like changes in the magnetization. These coincide with the HF crossings at $B_z^{(n)} \approx n \times 230$\,Gs ($n \in \mathbb{Z}$), measured optically (Fig. \ref{Fig:Long_Field_0_Trans}), where thermally activated resonant quantum tunneling\cite{Hartmann-Boutron:1996vo,Leuenberger2000,Giraud:2001vn} occurs. The steps can be rounded not only by raising the temperature, but also by slowing the longitudinal  field sweep (Fig.~\ref{Fig:Hyst_Sweep}), which provides a larger chance for thermal fluctuations to sample excited states, or by applying a transverse field. The latter enhances both the quantum tunneling (Fig.~\ref{Fig:Hyst_Trans}), also responsible for the avoided level crossings seen at equilibrium in our FTIR spectra, as well as the spin-phonon coupling, speeding up spin relaxation away from anti-crossings\cite{Malkin:2005qf,Bertaina:2006wd}. Fig. \ref{Fig:Spin_Relaxation_Hyst} illustrates the dramatic reduction in hysteresis due to a strong transverse field of 1\,T: the magnetization steps are removed in favour of a nearly smooth Brillouin-like function.  Fig. \ref{Fig:Spin_Relaxation} shows the quantum tunneling- and thermally-assisted approach to an equilibrium with positive magnetization along the trajectory indicated by the black arrow in Fig. \ref{Fig:Spin_Relaxation_Hyst}, where, after polarization by a strong longitudinal field $B_z$=-1000\,Gs and a subsequent sweep across $B_z$ = 0 to $B_z$ = 120\,Gs, the magnetization initially has a strongly negative value.

\subsection{Spectroscopic state mapping}
\begin{figure}[H]
	\centering
    \begin{tabular}{ccccc}
        \vspace{-3mm}
        \subfloat{
        \hspace{-4mm}
            \raisebox{40mm}{\stackinset{l}{80pt}{t}{5pt}{(a)}{\includegraphics[width =0.31\linewidth]
            {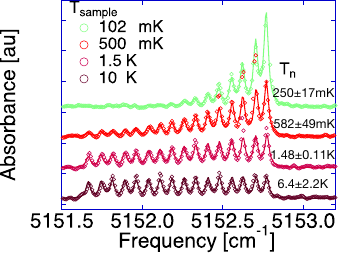}}
            }
            \label{Fig:115Gs_Spectra}
        }&
        \subfloat{
        \hspace{-60mm}
              \stackinset{l}{80pt}{t}{12pt}{(b)}{\includegraphics[width =0.345\linewidth,valign=b]{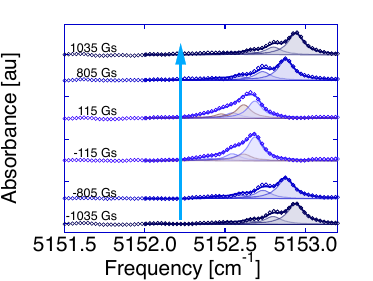}}
              \label{Fig:50mK_Fits}
        }&
        \hspace{-13mm}
        \subfloat{
            \stackinset{l}{75pt}{t}{9pt}{(c)}{\includegraphics[width =0.35\linewidth,valign=b]{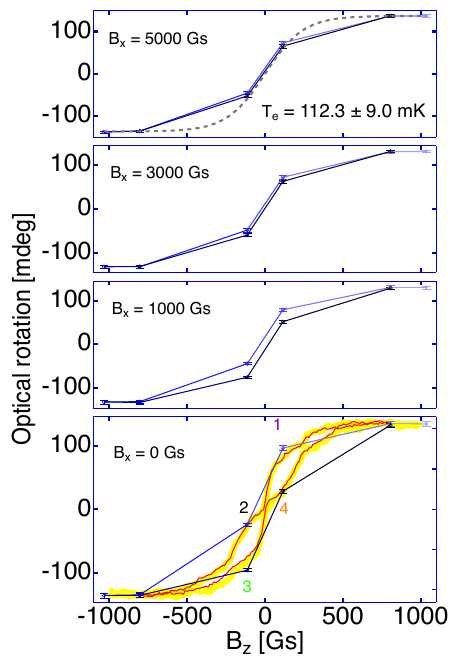}}
            \label{Fig:Rotation_Diff_Bx}
        }&
        \subfloat{
        \hspace{-9mm}
            \raisebox{24mm}{\stackinset{l}{85pt}{t}{13pt}{(d)}{\includegraphics[width =0.34\linewidth]
            {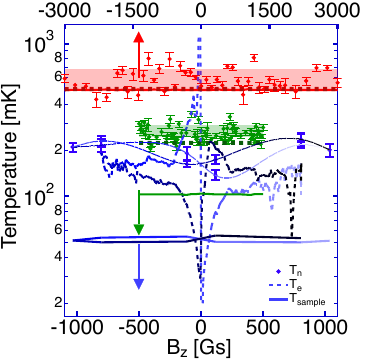}}}
            \label{Fig:T_n+T_e}
        }&
        \hspace{-65mm}
        \subfloat{
            \stackinset{l}{97pt}{t}{5pt}{(e)}{\includegraphics[width =0.34\linewidth,valign=b]{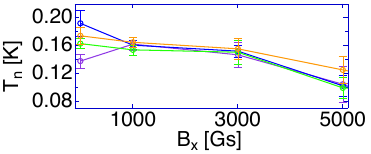}}
            \label{Fig:Tn_Vs_Bx}
        }
    \end{tabular}
    \caption{Thermalization. \protect\subref{Fig:115Gs_Spectra} FTIR spectra at different sample temperatures, with $B_z$=115\,Gs. On the right are $T_n$ values from fits to the HF lines. \protect\subref{Fig:50mK_Fits} FTIR spectra at the lowest temperature, and their fits, showing the constituent lines. The minority spins at $\pm$115\,Gs are shaded in red. The arrow shows the sweep direction. \protect\subref{Fig:Rotation_Diff_Bx} Closing of the optical rotation hysteresis loop at low temperature as the transverse field is increased. In the bottom panel the low-temperature magnetization, as obtained through a continuous sweep with FTIR source on (red), and the 120\,mK sweep from Fig. \ref{Fig:Hyst_Temp} (yellow highlight) are overlaid. The grey dashed line in the top panel is a paramagnetic fit. \protect\subref{Fig:T_n+T_e} Effective $T_n$ from fits to the HF spectra (markers), mean $T_e$ from the Faraday rotation (dashed lines) and sample temperature read by the temperature sensor (solid lines). The shaded areas indicate the average $T_n\,\pm$ standard deviation. The dashed blue line through the low-temperature $T_n$ is a guide to the eye and shows the sweep progression, from light to dark. The colors follow from the markers in Fig. \ref{Fig:Faraday_Magnetization}. \protect\subref{Fig:Tn_Vs_Bx} The decrease in $T_n$  at the 4 numbered points from Figure \protect\subref{Fig:Rotation_Diff_Bx} as a function of transverse field.}
	\label{Fig:Hi_Res_Faraday_FTIR}
\end{figure}

We combined  FTIR spectroscopy of high energy resolution (\!$\Delta\nu_\text{sp}\!\!
\ll\!\Delta\nu_\text{HF}$) with magnetization (Faraday effect) measurements for $B_z$ sweeps within $\pm$3000\,Gs. Here we use the transition to the lowest singlet in $^5I_7$ (excitation energy $5152\,{\rm cm}^{-1}$). Unlike the level we used for Fig. \ref{Fig:Introduction}, it is separated from other CF levels by more than the HF splitting of the ground doublet, which avoids the overlap of spectra pertaining to different electronic transitions. Fig. \ref{Fig:115Gs_Spectra} shows the effect of temperature on the spectra. For a nominal sample temperature of $T_\text{sample}$ = 10\,K, which corresponds to an energy of 0.86 meV and 6.95\,cm$^{-1}$, the 16 electronuclear levels with their net bandwidth of 2.2\,cm$^{-1}$ are nearly equally populated, and fitting the peak intensities using a Gibbs distribution can merely place a lower bound of $T$ on the electronuclear system temperature. As we lower $T_\text{sample}$ towards its base temperature, we see a continuous growth in the relative spectral weights for the higher photon energies, corresponding to increased occupancy of lower lying electro-nuclear levels in the ground state manifold. What is interesting, though, is that while the populations for $T_\text{sample}$=1.5\,K are described by effective temperatures very close to $T_\text{sample}$, for $T_\text{sample}$=0.5\,K and 102\,mK, the spectral peaks decay less with level energies than expected. The simultaneous net magnetization (Faraday) data in Fig. \ref{Fig:Faraday_Magnetization} exhibit a similar qualitative effect: the $B_z/T$ scaling expected for the magnetization of free ions is  violated for the two temperatures below 0.5\,K. This leads us to introduce an effective electron temperature $T_e$ for which $B_z/T_e$ scaling obtains, and so defines the horizontal axis of Fig. \ref{Fig:Faraday_Magnetization}. The Faraday effect only sees the electron spins, whereas the FTIR instrument sees all electronuclear levels, so we can use the values of $T_e$ to constrain fits of the FTIR peak intensities where nuclear spin temperatures $T_n$ are the free parameter. The upshot of the analysis of both Faraday and FTIR data is that as $T_{\rm sample}$ is lowered, there are growing deviations between the three effective temperatures. In particular, for $T_{\rm sample}=500$\,mK, Fig.~\ref{Fig:Faraday_Magnetization} (red traces) shows that the magnetization curve corresponds to an electron spin temperature $T_e=507\pm42$\,mK. Fitting the weights of the HF spectra to a Gibbs distribution yields a nuclear spin temperature of $T_n = 590\pm96$\,mK, indicating that both the electronic and the nuclear spins are in or close to thermal equilibrium with the phonons. When the temperature was allowed to reach its base level, which was read as $T_{\rm sample}=102$\,mK, the green traces show, however, elevated electron and nuclear spin temperatures of $T_e=225\pm6$\,mK and $T_n=261\pm 29$\,mK, respectively. It is plausible that the phonons inside the sample assume a similar elevated temperature, hotter than that in the bath outside, since the energy continuously injected by illumination and the field sweep cannot be evacuated quickly enough as opposed to the 160mK trace in Fig.~\ref{Fig:Hyst_Temp} where there is no FTIR illumination and the temperature is controlled. 

\subsection{Spectroscopic maps of the hysteresis}
To examine spectroscopically the magnetization steps seen in Fig. \ref{Fig:Hyst} we resorted to weaker FTIR illumination and to faster spectral measurements. We relaxed the instrument resolution relative to that for Fig.~\ref{Fig:115Gs_Spectra}, with resulting spectra exemplified in Fig. \ref{Fig:50mK_Fits} for a variety of longitudinal fields at base temperature. This enabled sweeps of $B_z$ at the relatively fast average rate of 400\,Gs/min over the range $\pm$1035\,Gs, which at the low $T_{\rm sample}\approx$50\,mK was sufficient for magnetization saturation. However, the rare earth ion system temperatures are likely to be higher than $T_{\rm sample}$\,. We estimate this effect, as shown in Fig.~\ref{Fig:Rotation_Diff_Bx}, bottom panel, by matching the optical rotation under FTIR illumination (blue in the main figure) with those from hysteresis loops (with no illumination) in Fig.~\ref{Fig:Hyst_Temp}. We find a good match with the non-illuminated trace at $T_{\rm sample}=$120\,mK, well in excess of the thermometer reading $\approx$50\,mK in the presence of illumination. This suggests that the phonons inside the sample are heated by the decay of excitations induced by illumination. The continuous field sweep with FTIR illumination (red line) passes through the discrete points (blue diamonds) where the FTIR spectra were measured. The fact that the Faraday rotation did not change during the spectroscopic measurement (lasting approximately 80s, during which the field sweeps were paused) demonstrates that the electron spin dynamics are very slow away from level crossings.    

In Fig.~\ref{Fig:50mK_Fits} we also plot resolution-broadened peaks representing separate electronuclear contributions to the spectra, including before and after the $B_z=0$ crossing. To our knowledge this is the first time the detailed populations, corresponding to the peak amplitudes,  on both arms of a hysteresis loop of a large spin magnet have been measured. Again we fit the measured populations to Gibbs distributions, the electronic spin orientation being constrained to match the measured magnetization, and supposing a common effective nuclear temperature $T_n$. Fig. \ref{Fig:T_n+T_e} shows the results of the fits, alongside those for the high resolution, higher temperature data of Fig. \ref{Fig:115Gs_Spectra}. Next to the nuclear temperatures (points), we plot $T_{\rm sample}$ (solid lines) and electron temperatures (dashed lines), the latter defined  simply as $T_e = \mu B_z/k_B \tanh^{-1}{M}$, where $M$ is the magnetization as measured by the Faraday rotation, nothwithstanding the absence of ergodicity in the electronic spin subsector needed to justify the notion of an actual electron temperature. As noted above, already at the two higher temperatures, the typical nuclear temperature is higher than the electron temperature, which in turn is higher than $T_{\rm sample}$. However, the truly dramatic result is divergence of the three temperatures on cooling to the base temperature of the cryostat. In addition, at the lowest temperature we see hysteresis in all three, including cooling of not only the electrons but also the sample thermometer via adiabatic demagnetization. The accelerated spin relaxation at the level crossings tends to align $T_e$ and $T_n$, especially on traversal of $B_z= 230Gs$, both being elevated due to the energy injected by the field sweep that outpaces the relaxation away from the level crossings. 
In the regime $B_z>0$ the electron spins relax mostly by thermal bath-induced flips. This tends to decrease the effective $T_e$ toward $T_{\rm ph}$ while the spin flip against the orientation of the nuclear spin tends to increase the energy stored in the latter, thus pushing $T_n$ up. This is observed in the experimental data (blue markers tracing $T_n$) in Fig.~\ref{Fig:T_n+T_e}. From these considerations the lattice (bath), electron and nuclear temperatures are expected to follow the ascending order  $T_{\rm ph}< T_e < T_n$ during the sweep.  
All temperatures will eventually equilibrate to the same value, at a rate limited by the electronic spin flips (at moderate $B_z$) or the even slower nuclear spin-phonon relaxation in the presence of strong electron spin polarization, which is estimated to last for $O(10^3)$ seconds, cf.~\ref{Sec:n_phononcoupling}.  

Table \ref{tab:T_M} provides a summary of magnetizations and $T_n$ for each measurement point along the hysteresis curves (blue markers in Fig. \ref{Fig:Rotation_Diff_Bx}, bottom panel). 

We now turn our attention to the effect of a transverse field on $T_e$ and $T_n$ at low temperature. Fig. \ref{Fig:Rotation_Diff_Bx} shows a closing of the magnetization hysteresis with increasing transverse field due to the enhanced spin-lattice relaxation rates. This is similar to Fig. \ref{Fig:Hyst_Trans} but now in the  presence of the FTIR light source. At a transverse field of 5000\,Gs we fit to a paramagnetic curve to obtain an equilibrium $T_e$ of 112.3$\pm$9.0\,mK. In Fig. \ref{Fig:Tn_Vs_Bx} we see a convergence and lowering of $T_n$ with transverse field, with $T_n=108.4\pm74.1$\,mK at $B_x=5000$\,Gs.
 
\subsection{Modeling of dynamics}

\noindent Our magnetization measurements show the well-known steps characteristic of single molecule magnets. In addition, the FTIR data show directly that heightened nuclear spin excitations accompany the magnetization steps. This corresponds to enhanced effective nuclear spin temperatures, which are even found at somewhat elevated cryostat temperatures where there are no magnetization steps, while $B_z/T_\text{e}$ scaling, with $T_e>T_\text{sample}$, 
indicates elevated effective electron spin temperature. 

\begin{figure}[!ht]
	\centering
 	{\subfloat{
		\stackinset{l}{-20pt}{t}{0pt}{(a)}{\includegraphics[width = 0.6\linewidth,valign=t]{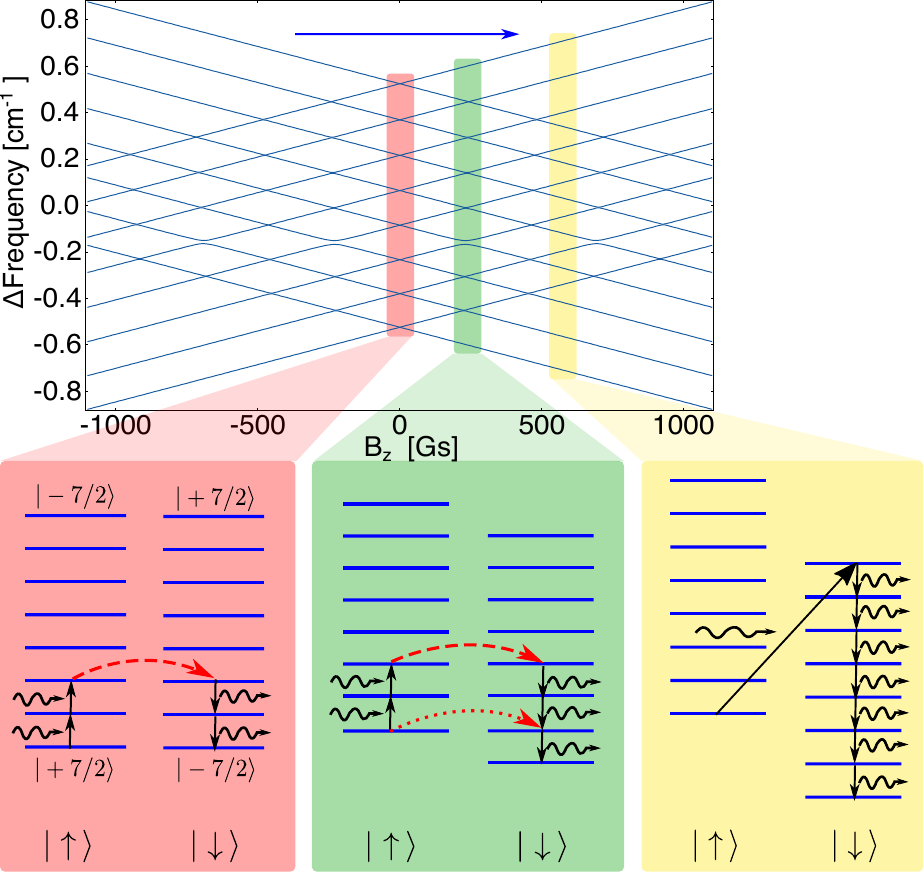}}
		\label{Fig:relaxation_pathways}
	}}
    \caption{\protect\subref{Fig:relaxation_pathways} Relaxation paths of the electronic moments. At $B_z=0$ (red) the dominant relaxation path includes two excitation steps, taking the nuclear spin to $I_z = 3/2$, where the finite gap allows for resonant quantum tunneling of the combined electronuclear degrees of freedom. Subsequently the ion relaxes as the nuclear spin excitation diffuses to a other ions. At $B_z= \pm 230 \text{\,Gs}$ (green) the dominant channel is again via absorption of two hyperfine excitations, direct Landau-Zener tunneling being much weaker. Between crossings of the HF levels (yellow), relaxation typically occurs through a non-resonant flip of the electronic spin involving phonons. The blue arrow indicates the sweep direction.}
	\label{Fig:Spi_Flip}
\end{figure}

The sudden magnetization steps can be understood quantitatively within the 
relaxation model of Fig. \ref{Fig:Spi_Flip}. It describes the thermally assisted resonant tunneling of the electron spin at $B_z=0$ and $B_z= \pm 230 \text{\,Gs}$ between two electronuclear states with flipped electron spins. 

Similar to Refs.~[\,\onlinecite{Leuenberger2000}, \onlinecite{Leuenberger2000a}], we model the activated tunneling using a Lindblad master equation, see Methods, Eq.~\ref{Eq:Rate_Eq}. For the moment we leave the nature of the activating bath open, assuming it to be characterized by an inverse temperature $\beta$ and an activation attempt rate $\kappa$. The transition rates between electronuclear states, and thus the thermalization rate, depend sensitively on the degree of hybridization of HF states and thus on the longitudinal field $B_z$. Each of the HF crossings allows the electronuclear complex to tunnel and flip the electron spin, adding up to the total spin-flip rate $\Gamma_\mathrm{eff}(B_z) = \sum_{i=1}^7 \Gamma_i(B_z)$. For the crossing at $B_z=0$ and in small applied transverse fields $B_x$, where the tunnel gaps are much smaller than the nuclear temperature, $\Delta_i \ll T$, we find
\begin{equation}
	\Gamma_{i}(B_z) = \frac{\Delta_i^2}{(2\mu B_z)^2 + \Delta_i^2} (e^{-\beta \epsilon_i} \kappa \,m_i).
\label{Eq:Gamma_ֵEff_Main}
\end{equation}
Here, $\Delta_i$ are the gaps of the different HF crossings ($i=1$ being the first excited crossing), $\kappa$ is the diffusion (decay) rate of the first excited HF state ($I_z =\pm 5/2$) at $T=0$ and $\epsilon_i \approx i \times E_\mathrm{HF} = i \times 0.21\text{\,K}$ is the activation energy to the $i$-th state. $m_i$ are ratios of matrix elements ($O(1)$) of the nuclear spin raising operator between the different HF states, see Methods Sec.~\ref{Sec:Activated_Tunneling} for more details. The Lorentzian factor in Eq.~\eqref{Eq:Gamma_ֵEff_Main} describes the probability that the bath-induced transition from the (little hybridized) state $i-1$ to the more strongly hybridized state $i$ and its subsequent decay of the excitation flips the electron spin. This process is operative in the field window $|\mu B_z| \lesssim \Delta_i $, where the hybridization of the $i$'th HF level is substantial. The second factor in Eq.~\eqref{Eq:Gamma_ֵEff_Main} is the attempt rate of the tunneling process. Note that the direct crossing ($i=0$) contributes negligibly to relaxation at our experimental temperatures. Its hybridization gap $\Delta_0$ is insignificant as it involves high-order tunneling ($\lvert\Delta I_z\rvert=7$).

\noindent In the dilute limit, the probability for an electron spin to flip when sweeping through \mbox{$B_z=0$} is found by integrating over the spin-flip rate,
\begin{equation}
	P_\mathrm{flip} \approx   \frac{1}{2} \left(1- e^ {- \int_{\infty}^{\infty} dt\, \Gamma_\mathrm{eff}(B_z(t)) } \right)
	\approx \frac{1}{2} \left(1- e^ {-\frac{\pi}{2} \sum_{i=1}^7 \frac{\Delta_i}{v} e^{-\beta \epsilon_i} \kappa m_i} \right),
\label{Eq:P_0Gs}
\end{equation}
which can be understood as a Landau-Zener probability\cite{Zener1932} of parallel flip channels, whereby  $v= |\mu \, \mathrm{d}B_z/\mathrm{d}t|$ denotes the rate of change of the Zeeman energy. The dominant activated tunneling channel $i$ depends on the numerical factors in Eq.~\eqref{Eq:P_0Gs}. At the base temperature of our experiments and at small external fields $B_x$, the $i=2$ ($I_z =\pm 3/2$) channel dominates, as sketched in Fig.~\ref{Fig:relaxation_pathways}.

\noindent Let us next discuss the spin-flip probability at $B_z=0$ as a function of the applied transverse field at longitudinal sweep rate. At large external transverse fields $B_x$, the excited gaps become of order of the temperature, $\Delta_i \gtrsim T$. In this regime the theory of Refs.~[\,\onlinecite{Leuenberger2000,Leuenberger2000a}] is no longer valid, and the equilibration can flip more than 50\% of the spins, see Methods Eqs.~\eqref{Eq:P_Flip_Exact} and \eqref{Eq:P_Flip_Approx}. If the activation were by phonons we may approximate the relevant bath temperature by the thermometer reading right after crossing $B_z=0$, using the (discrete) measurements shown in Fig. \ref{Fig:T_vs_B}). From this we would extract for the only free fit parameter an attempt rate of $\kappa = 57\pm 4 \text{\,kHz}$. 
However, by estimating the coupling of nuclear spins to acoustic phonons, using the electron-phonon coupling determined in Ref.~\citeonline{Bertaina:2006wd} we find the actual phonon rate to be 7 orders of magnitude smaller than this value (cf. \ref{Sec:n_phononcoupling}). Even by assuming a much higher phonon temperature a large discrepancy in the attempt rate would persist, ruling out phonons as the relevant bath. We identify a much more efficient activation channel instead: the diffusion of nuclear spin excitations among Ho ions, mediated by hyperfine interactions and dipolar coupling between the virtual electron excitations they create. Swaps of hyperfine states between ions with nuclear projections $I_z$ and $I_z+1$ are indeed nearly resonant processes. The remaining random energy mismatches due to internal dipolar fields and variations of the Ho hyperfine field can be bridged by energy exchange with the bath of fluorine spins (thereby avoiding the extremely weak coupling to phonons). We estimate the resulting hyperfine diffusion rate to be of order $O(100 Hz)$, five orders of magnitude faster than activation rates by phonons. The remaining discrepancy with the fitted $\kappa$ can be resolved at a qualitative level by noticing that the constant field sweep drives the nuclear spin temperature to significantly higher values than the temperature read from the thermostat outside the sample. The larger Boltzmann factors compensate for the lower attempt rate, rationalizing the observed magnetization drops. A fully quantitative theory is however beyond the scope of this paper.

\noindent At $B_z = \pm 230\text{\,Gs}$ (after crossing $B_z=0$) we find the crossings between the excited states $\ket{\uparrow, -3/2}$ and $\ket{\downarrow, 1/2}$ (or their time-reversed counterparts at $B_z=-230 \text{\,Gs}$) to be avoided by large gaps, as sketched in Fig.~\ref{Fig:relaxation_pathways}. These persist even in the absence of an external transverse field. In principle, the activated tunneling can be treated in a similar way to $B_z=0$, see Methods section~\ref{Sec:Activated_Tunneling}. This gap and the tunneling are so strong, that one expects the spin dynamics to reach complete equilibrium at $B_z=\pm 230$\,Gs. However, this seems in direct contradiction to the experiment, where the system settles in a less magnetized state than expected at the external bath temperature. We attribute this to a phonon bottleneck effect\cite{Scott1962,Abragam1976}. The spins of the Ho subsystem equilibrate among each other essentially without the involvement of phonons, resulting in an effective $T_n=T_e$, substantially above the phonon temperature. To cool down, the spins must undergo transitions emitting phonons that escape the sample.
However, the number of  relevant low energy phonons  is many orders of magnitude smaller than the number of Ho ions. Their ratio $N_\mathrm{ph}/N_\mathrm{Ho}$,  estimated\cite{Barbara:1999fk} as \mbox{$3.0 \times 10^{-7}$}, controls the ratio between the energy relaxation rate of the spin subsystem and the typical escape rate ($\kappa_{\rm ph}\sim c_s/L =O( 1\mu s)$, where $c_s$ is the sound velocity and $L$ is a typical sample dimension) of phonons from the sample. 
For most sweep rates the equilibration with the external bath takes much longer than the time to pass through the level crossing. The Ho spins then equilibrate to an effective temperature corresponding to the energy in the spin system before the crossing. 
The phonon bottleneck that limits relaxation at level crossings, and the generally slow spin-phonon relaxation elsewhere, are thus seen to be at the heart of hysteresis effects. Note that the phonon bottleneck is irrelevant at $B_z=0$, where spin flips do not dump excess energy into the lattice. The experimental observation of this phonon bottleneck is important in the context of low-temperature phenomena in this and similar materials. It likely plays a role in the emergence of out-of-equilibrium states in lightly doped materials, especially with weak heat bath contact, which display hole burning at ultra-low frequencies and entail magnetic response and specific heat features opposite to those in a fully relaxed dipolar spin glass state.

\subsection{Summary and Significance}
Our combined optical spectroscopy and magnetometry provide an unprecedentedly detailed portrait of a prototypical "single atom" magnet, including especially the population dynamics underlying the magnetization steps also displayed by single molecule magnets. Quantum fluctuations, conveniently regulated via externally imposed transverse fields, influence the dynamics especially at electronuclear level crossings encountered in longitudinal field ($B_z$) sweeps. Our spectroscopic maps reveal and confirm the three different classes of crossings and their dependence on the transverse magnetic field. 
While one might think that at 0.3\% rare earth site occupancy the interactions among Ho ions are negligible, we find evidence to the contrary: The thermally assisted quantum tunneling cannot be explained unless the dipolar-induced diffusion of hyperfine excitations is invoked for the thermal activation.   
We use this knowledge and a single nuclear spin diffusion rate $\kappa$ to account quantitatively for the impact of quantum and thermal fluctuations on the electronuclear level populations as the longitudinal field is swept. At lower temperatures the nuclear spins can still equilibrate amongst each other through multi-ion flip processes that do not involve any phonons, and thus can establish and maintain a nuclear spin temperature much higher than the phonon temperature. The latter is very long lived at low $T$ where both electron and nuclear spin-lattice relaxation are strongly suppressed,  particularly so in large fields $B_z$. At level crossings where the spin dynamics are exceptionally fast,  equilibration with the bath is instead hindered by the phonon bottleneck, so that the sample only equilibrates internally to an elevated temperature.

Our results are significant for several reasons. First, they show that even free space optical spectroscopy at temperatures of order 0.1\,K is possible in a relatively commonplace "dry" dilution refrigerator modified with appropriate windows. An ability to perform spectroscopy and see sharp lines implies also the possibility of optical control, meaning that for future quantum science and technology, multiplexed wires and  RF lines can be avoided as samples and devices with multiple active regions, corresponding \textit{e.g.} to qubits, are introduced. Second, our data for a dilute rare earth salt suggest that phonons can actually be irrelevant for thermalization, implying that such salts can be solid state qubit hosts which can be most productively thought of as ion traps with symmetry defined by crystal structure. As for non-interacting ions, there is a decoupling of electron and nuclear spins at low temperatures,  while for concentrated samples (pure LiHoF$_4$) where the rare earth ions interact, we see quantum dynamics well described by a mean field approach where electrons and nuclei are fully equilibrated \cite{Ronnow2007, Stamp2024}. This dichotomy suggests a  cross-over between the two regimes, which will depend on the density $x$ of rare earth ions. $x$ controls, among others, the magnitude of internal transverse fields and thus the quantum tunneling. Indeed, at intermediate concentrations, notably $x=0.044$,  there appears to be an approach to an "ordinary" spin glass or an unconventional state ("antiglass")\cite{Wu:1991} resembling more a valence bond liquid \cite{Ghosh2003} and with a remarkable capacity for narrow bandwidth hole-burning \cite{Ghosh2002,Silevitch2019}, depending on whether the coupling to the thermal bath is strong or weak, respectively \cite{Schmidt:2014}. We therefore suggest a programme of optical diagnostics and theory to follow the evolution from paramagnet to "antiglass" and then spin glass and ferromagnet as a function also of coupling to the bath.    
\begin{methods}
\section{Experimental}
\label{sec:Experimental}
We mounted the \LiHoF\:sample (2$\times$5$\times$5\,mm$^3$, $a$-axis along the 2\,mm side) in an Oxford Instruments Triton 200 dilution refrigerator using silver paste for thermal conduction. The refrigerator was equipped with Z-cut sapphire windows in four directions. The light source of the FTIR spectrometer (Bruker IFS125HR for the high-resolution measurements, Bruker Vertex 80v for the spectroscopic hysteresis measurements) was a halogen lamp. The output of the FTIR was polarized with a calcite polarizer, directed along the $c$-axis of the crystal and detected with a nitrogen-cooled MCT detector. The signal was fed back into the FTIR for processing. We collected reference spectra by bypassing the refrigerator. A CW laser at 670\,nm was directed along the $c$-axis of the crystal and detected using a balanced photodiode bridge. The two photodiode signals were recorded with a DAQ card and converted to a rotation angle. The refrigerator was equipped with a vector magnet capable of generating 2\,T in any direction. The magnet axes were aligned to the sample orientation by minimizing the magnetization at transverse fields. To minimize heating, the laser power was attenuated to 30\,nW ($\approx$1\,$\mu$W/cm$^2$) and the FTIR output onto the sample was filtered with a germanium window.

\noindent For the Faraday measurements $B_z$ was swept between $\pm1035$\,Gs at different fixed values for $B_x$, $T$ and $dB_z/dt$. The window facing the FTIR was blanked for these measurements. For the spectrally resolved hysteresis measurements both the Faraday laser and the FTIR light traversed the sample. $B_z$ was swept at 400\,Gs/min between $\pm1035$\,Gs, stopping at $\pm805$\,Gs and $\pm115$\,Gs to measure low-resolution ($\Delta \nu=0.075$\,cm$^{-1}$) spectra. These magnetic field stops were chosen so as to be close to the $B_z=0$ crossing and at full magnetization, while far away from crossing/anti-crossings of HF levels. The sweep cycle was repeated 10 times. The population of the individual HF states was found by integrating Lorentzians fitted to the experimentally measured peaks. We assumed that all peaks have the same linewidth for each magnetic field ($0.123\pm0.007\,\text{cm}^{-1}$ on average) and that the HF spacing was uniform for each magnetic field ($0.147\pm0.008\,\text{cm}^{-1}$ on average).

\noindent To measure the spin relaxation time between crossing points the longitudinal field was swept to -1000\,Gs, then to 120\,Gs where the field was kept constant and the rotation was measured over time at $T_\text{sample}<40\,\text{mK}$.

\section{Spectrum calculation
\label{Sec:Spectrum}}

\subsection{CF Hamiltonian}
The  Hamiltonian of a single Ho ion in Ho$^{3+}$:LiYF$_4$ is 
given by 
\begin{equation}
\begin{split}
	H(J) =& H_\mathrm{CF} + H_Z + H_\mathrm{HF} \\
	=& \left( \sum_{l=2,4,6} B_l^0 O_l^0(\vec{J}) + \sum_{l=4,6} \sum_{m= \pm 4} B_l^m O_l^m(\vec{J}) \right) \\
	&+ g_J \mu_\mathrm{B} \vec{B}\cdot\vec{J} + A_J \vec{J}\cdot \vec{I}.
\end{split}
\label{Eq:H_Single_Ion}
\end{equation}
where the crystal field (CF) Hamiltonian of the ground state $J=8$-manifold is expressed with Stevens operators  $O_l^m(\vec{J})$~\cite{Stevens1952,Hutchings1964} consistent with $S_4$ symmetry, with coefficients taken from Ref.~\onlinecite{Beckert:2022gb}. 
The second and the last term describe Zeeman and HF interactions 
 in terms of the electronic angular momentum operator $J$,
$g_J=5/4$ being the Land\'e $g$-factor and $A_J= 39 \text{\,mK}$  the HF coupling in the ground-state manifold~\cite{Beckert:2022gb}), with $I=7/2$ the Ho nuclear spin. 
The total magnetic field $\vec{B}$ is composed of the external field and internal dipolar fields, mostly from other Ho ions.

\subsection{CF and HF states}

The CF eigenstates are classified by the irreducible representations of $S_4$ (singlets $\Gamma_1,\Gamma_2$ or doublets $\Gamma_{3,4}$. The ground state is an Ising doublet, for which we choose the standard basis $\ket{\uparrow},\ket{\downarrow}$ with maximal magnetic moment $\pm \mu= \pm m_z g_L \mu_B$ along the crystalline $c$-axis, where $m_z=5.4$. The HF coupling introduces an eightfold splitting of the ground state (see Fig.\ref{Fig:Long_Field_0_Trans}). We label the HF eigenstates states by $\ket{\uparrow/\downarrow, I_z} \equiv \ket{\uparrow/\downarrow} \otimes \ket{I_z}$, $I_z$ being the dominant projection of the nuclear spin on the $z$ axis (with small admixtures of other electronuclear configurations due to the transverse HF interactions).

\subsection{Crossing of HF states in longitudinal magnetic fields}

A longitudinal field $B_z$ lifts the degeneracy of the electronic doublet and allows to drive crossings between HF states having opposite magnetic moments. The two states $\ket{\uparrow, I_z^\uparrow}$, $\ket{\downarrow, I_z^\downarrow}$ undergo a crossing  at 
\begin{equation}
	B_{n=-(I_z^\uparrow + I_z^\downarrow)} \approx n \times A /(2 g_L \mu_B)  \approx n\times 230 \text{\,Gs},
\label{Eq:Bz_Crossing}
\end{equation}
which is proportional to the integer $n=-(I_z^\uparrow + I_z^\downarrow)$. Since $B_z$ does not break the $S_4$ symmetry, most of these crossings are protected (not avoided) as they occur between states of different $S_4$ symmetry. Only crossings of HF states that share the same irreducible representation are avoided in a purely longitudinal field. These correspond to pairs of states $\ket{\uparrow, I_z^\uparrow}$ and $\ket{\downarrow, I_z^\downarrow}$ for which the nuclear spins differ by \mbox{$\Delta I_z \equiv I_z^\downarrow - I_z^\uparrow  =  2 \, \mod \, 4$} .

\subsection{Crossing of HF states in transverse magnetic fields}

In contrast, a transverse field $B_x$ breaks both time-reversal and the $S_4$ symmetry, and thereby lifts all degeneracies. Thus, a small transverse field opens gaps at all HF crossings that are protected in a purely longitudinal field. The $S_4$ symmetry dictates that level crossings with odd $\Delta I_z  \in \{1,3\} \, \mod \, 4$ are avoided by a gap linear in small $B_x$, while the gap of avoided crossings with even \mbox{$\Delta I_z \equiv 0 \, \mod \, 4$} are quadratic in small $B_x$. The coefficients of this leading $B_x$ dependence are derived in detail in Sec. \ref{Sec:Perturbation_Theory}.

\section{Thermally activated quantum tunneling}
Here we develop the theory of thermally activated quantum tunneling in 
\LiHoF\ close to $B_z$-tuned avoided HF crossings. We make three simplifying assumptions: i) We neglect the internal longitudinal dipolar fields, which are small given the small concentration of Ho.   
ii) we describe the coupling strength of hyperfine transitions to the activating bath (be it phonons or a reservoir of HF excitations on other Ho ions) by  the rate $\kappa$ 
with which the first nuclear spin excitation decays back to the ground state. iii) We assume the activating reservoir to be in equilibrium at an effective temperature $T= \beta^{-1}$. 
From the subsequent analysis we will conclude that the phonon bath is many orders of magnitude less efficient than diffusing nuclear spin excitations. For the latter we justify assumption (iii) in Sec.~\ref{Sec:Ho_Dynamics} (with a temperature $T=T_n$ that tends to lie above the phonon temperature),
while the diffusion rate is estimated in Sec.~\ref{Sec:Ho_Dynamics} to be of the order of  $\kappa=O(100 Hz)$. 
As for assumption (i), it can be dropped. The dipolar interactions will reset the local field of ions close to a flipping Ho spin. On average this enhances the time a given ion spends in the critical window where the tunneling dynamics is strongly enhance and thus increases the spin flip probability. We have verified that the effect is, however, modest.

\subsection{Theory of thermally activated quantum tunneling \label{Sec:Activated_Tunneling}}
We consider (single) Ho ions initialized in a longitudinal field $B_z$ at low temperatures, subject to a linear field sweep, $B_z(t)-B_z(0) \propto t$, characterized by the rate of increase of the Zeeman energy,
\begin{equation}
	v \equiv \left\lvert \frac{\mathrm{d}(\mu B_z(t))}{\mathrm{d}t} \right\lvert.
\end{equation}
For simplicity we restrict the discussion to sweeps from negative to positive $B_z$. 

\noindent Dipolar interactions between the Ho ions create (inhomogeneous) internal fields. Their typical value $B_{x,\mathrm{dip}}$ is given by the half-width-at-half-maximum of their distribution~\cite{Kittel:1953sc}, $$B_{x,\mathrm{dip}} \approx x \times 0.77\text{\,T.}$$ At a Ho dilution of $x=0.3\%$, one has $B_{x,\mathrm{dip}} \approx 23$\,Gs. 
The transverse components of these fields allow for spin flips in a vanishing (external) longitudinal field.

\subsection{Lindblad  equation}
\label{Sec:Lindblad}

We model the coupling to an activating bath by a Lindblad equation for the  density matrix $\rho$ of a single Ho ion~\cite{Albash2012},
\begin{equation}
\begin{split}
	\dot \rho(t) =& -\frac{i}{\hbar} [H(t), \rho(t)] + \sum_{n,m} \gamma_{nm}(t) \\
	&\times \left[ L_{nm}(t) \rho(t) L_{nm}^\dagger(t) - \frac{1}{2} \{L_{nm}(t)^\dagger L_{nm}(t), \rho(t)\} \right],
\end{split}
\label{Eq:Lindblad}
\end{equation}
where the operator $L_{nm}(t) = \ket{\psi_m(t)} \bra{\psi_n(t)}$ induces a transition from the instantaneous eigenstate $\psi_n(t)$ to $\psi_m(t)$. 
For the case of activation via nuclear excitation diffusion the transition rates $\gamma_{nm}(t)$ are given by 
\begin{equation}
\gamma_{nm} = \kappa\, p(m)  \frac{|\langle I_m| I_{+}+I_-|I_n \rangle|^2}{|\langle I_z=-7/2| I_-|I_z=-5/2 \rangle|^2},
\label{Eq:gamma_nm}
\end{equation}
where $p(m)$
is the probability of a neighboring Ho ion to be in nuclear spin state $I_m$ and have the same electronic spin projection as state $n$; $\kappa$ is the $T=0$ decay rate of the first excited hyperfine state in a polarized sample. Typically such hopping processes are only resonant if $I_n-I_m = \pm 1$ and, if the magnetic field is non-zero, if the two ions have the same electronic spin projection\footnote{For phonon assisted processes a phonon occupation number  replaces $p(I_m)$}.
No matter what the microscopic origin of the transitions from state $n$ to $m$, the rates obey the detailed balance condition,
\begin{equation}
	\frac{\gamma_{nm}}{\gamma_{mn}} = e^{\beta (\omega_n-\omega_m)},
\label{Eq:Detailed_Balance}
\end{equation}
such that at constant external field $B_z$ the master equation~\eqref{Eq:Lindblad} leads to thermal equilibrium. 

\noindent Following Refs.~[\,\onlinecite{Leuenberger2000,Leuenberger2000a}] we will reduce the Lindblad  equation to a classical master equation for the population of the instantaneous eigenstates. The main difference is that, because of large hybridization gaps and the comparatively slow driving, we need to retain the hybridization of HF states close to the crossing. This will modify the spin flip probability in large applied transverse fields.

\subsubsection{Level crossing at $B_z \approx 0$
\label{Sec:Tunneling_Bz=0}}

At \mbox{$B_z = 0$}, the time-reversed states $\ket{\uparrow,I_z}$, $\ket{\downarrow,-I_z}$ undergo a crossing which is avoided because of transverse fields. We approximate the Hamiltonian as eight blocks of two crossing HF levels, $H(t) \equiv \bigoplus_{i=0}^7  H_{i}(t)$, where
\begin{equation}
	H_{i}(t) = \epsilon_i+ \epsilon_B(t) \sigma^z + \frac{\Delta_i}{2} \sigma^x,
\label{Eq:H0}
\end{equation}
with linearly increasing bias \mbox{$\epsilon_B(t) \equiv v t$} and the Pauli matrices $\sigma^x,\sigma^z$. The gaps of the avoided crossings are denoted as \mbox{$\Delta_i\equiv \Delta \varepsilon (B_z=0, \Delta I_z= 7-2i)$} (see Sec.~\ref{Sec:Perturbation_Theory}). 
The index $i$ labels the eight pairs of crossing HF states  $\ket{\uparrow,-7/2+i},\ket{\downarrow,7/2-i}$. The mean of the two HF energies of a pair is $\epsilon_i \approx i\times E_\mathrm{HF}$. We will use the instantaneous eigenstates and energies of this Hamiltonian to calculate phonon transition rates. We denote them by
\begin{equation}
\begin{split}
	\ket{i,+} &= \phantom{-} \cos(\theta_i/2) \ket{\uparrow,-7/2+i} + \sin(\theta_i/2) \ket{\downarrow,7/2-i} ,\\
	\ket{i,-} &= -\sin(\theta_i/2) \ket{\uparrow,-7/2+i} + \cos(\theta_i/2) \ket{\downarrow,7/2-i}.
\end{split}
\label{Eq:Psi_H0}
\end{equation}
Their energies are
\begin{equation}
	\omega_{i, \pm} = \epsilon_i \pm \sqrt{\epsilon_B^2 + \left(\frac{\Delta_i}{2} \right)^2},
\end{equation}
with $\theta_{i}$ the root of $\tan \theta_i = \Delta_i/(2\epsilon_B)$, $\theta_i \in [0,\pi]$, omitting the time dependence of $\epsilon_B(t)$ for brevity. The states $\ket{i,+},\ket{i,-}$ correspond to the upper and lower energy branches, respectively.

\noindent \textbf{Choice of basis.} We will express the density matrix $\rho$ in Eq.~\eqref{Eq:Lindblad} in terms of suitable basis states. If $\Delta_i^2\gg (\hbar v)$, the eigenstates in sector $i$ evolve adiabatically.~\cite{Zener1932} The states $i=2$ and $i=3$ that constitute bottlenecks for the thermally activated tunneling, evolve adiabatically under our experimental sweep rates, even in the absence of an external transverse field. Under these adiabatic conditions, the diagonal and off-diagonal matrix elements of $\rho$ with respect to the instantaneous eigenbasis decouple in Eq.~\eqref{Eq:Lindblad}. This simplifies their equations of motion. On the other hand, under diabatic basis conditions, one better uses the unhybridized states as a basis. The lowest HF states, $i=0$, cross $B_z=0$ diabatically for all applied transverse fields, $\Delta_i^2\ll (\hbar v)$. We thus approximate $\Delta_0 = 0$, using the diabatic basis states $\ket{i=0,\uparrow}\equiv \ket{\uparrow,-7/2}$ and $\ket{i=0,\downarrow} \equiv \ket{\downarrow,7/2}$. Since the intermediate HF states $i=1$ only act as transition levels to reach the levels $i=2$ or $i=3$ that mediate quantum tunneling, it is immaterial for our formalism whether or not we retain the hybridization in that sector. We work below with instantaneous eigenstates and treating them as evolving adiabatically, \textit{i.e.}\ we neglect couplings of diagonal and off-diagonal terms. The states $i>3$ are irrelevant for spin-flips and will be completely neglected.

\noindent \textbf{Rate equation.}  We now reduce the Lindblad equation to a classical master equation for the diagonal elements $\rho_n \equiv \braket{n|\rho|n}$ in the appropriate basis. The coupled rate equations read
\begin{equation}
	\dot \rho_n = -\rho_n \sum_{m\neq n} \gamma_{nm}(t) + \sum_{m\neq n} \rho_m \gamma_{mn}(t).
\label{Eq:Rate_Eq}
\end{equation}
Note that the hybridizations (\ref{Eq:Psi_H0}) allow for finite transition rates $\gamma_{nm}(t)$ between states that differ in the dominant polarization of the electronic spin, thus enabling its flipping.

\noindent Since we only retain transitions that change the nuclear spin by $\Delta I_z = \pm 1$, the Lindblad operators in the master equation~\eqref{Eq:Rate_Eq} only connect sectors $(i,i+1)$, see Fig.~\ref{Fig:rates}. We can rewrite the evolution for states $\ket{i,\tau}$ with $\tau \in \{+,-\}$ (for $i>0$) explicitly as
\begin{equation}
\begin{split}
	\dot \rho_{(i,\tau)} =& -\rho_{(i,\tau)} \sum_{\sigma=\pm} \sum_{s=\pm1}  \gamma_{(i,\tau),(i+s,\sigma)}(t)  \\
	& + \sum_{\sigma=\pm} \sum_{s=\pm1}  \rho_{(i+s, \sigma)} \gamma_{(i+s, \sigma),(i,\tau)}(t) .
\end{split}
\label{Eq:Rate_Eq_2}
\end{equation}
Note that for the lowest states, $i=0$, we use the diabatic basis for which we retain the labels $\tau \in \{ \uparrow, \downarrow\}$.
\begin{figure}[ht]
	\centering
	\includegraphics[width = 0.48 \textwidth] {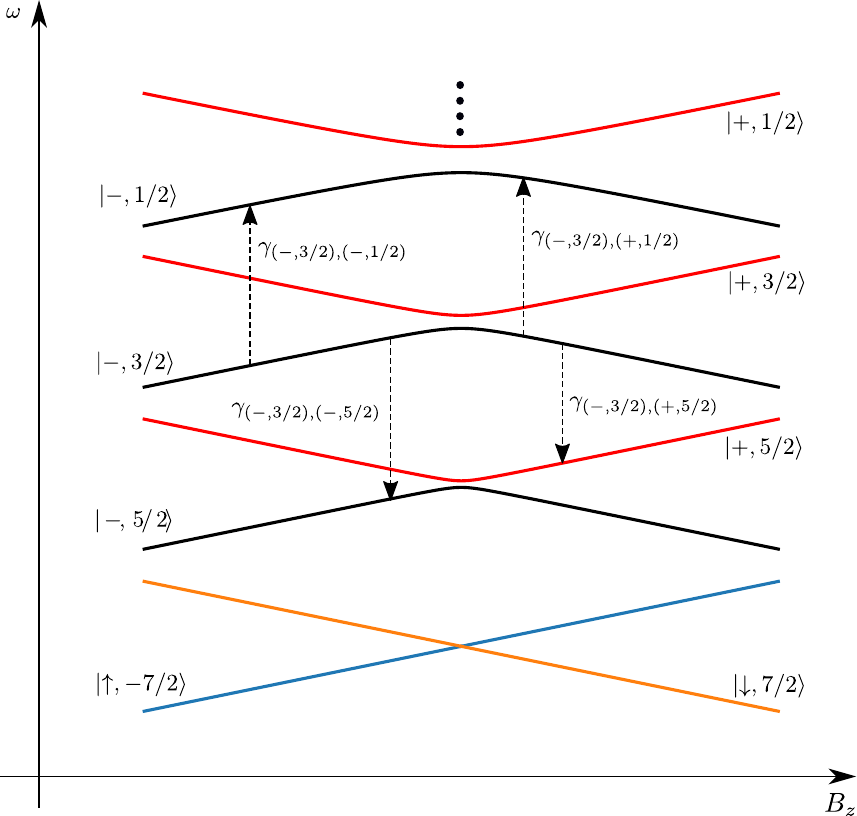}
	\caption{Sketch of the instantaneous eigenenergies of Hamiltonian~\eqref{Eq:H0} and the transitions in the rate equation~\eqref{Eq:Rate_Eq_2}. For readability, we show only the transitions starting at state $\ket{-,3/2}$, each plotted at different $B_z$. The rates are time-dependent via the (instantaneous) transition energies and matrix elements between eigenstates. \label{Fig:rates}}
\end{figure}

\noindent \textbf{Solution of the rate equations.} To solve the time-dependent rate equations~\eqref{Eq:Rate_Eq_2}, we proceed in the same fashion as Ref.~[\,\onlinecite{Leuenberger2000}]. At each $B_z(t)$ we assume that a stationary flow establishes,
\begin{equation}
\begin{split}
	\dot \rho_{(0,\uparrow)} &= -J(t), \\
	\dot \rho_{(0,\downarrow)} &= \phantom{-} J(t), \\
	\dot \rho_{(i,\pm)} &= \phantom{-} 0 \ (\forall i \geq 1),
\end{split}
\label{Eq:Prob_Current}
\end{equation}
with the time-dependent probability current $J(t)$. This approximation assumes that part of the population of $\ket{0,\uparrow}$ is transferred to $\ket{0,\downarrow}$, while the in- and outflow of the excited populations $i\geq 1$ compensate each other.

\noindent At any instantaneous field $B_z(t)$, we can solve Eqs.~\eqref{Eq:Rate_Eq_2},\eqref{Eq:Prob_Current} to obtain the current $J(t)$. The structure of the equations imply that the current  takes the form
\begin{equation}
	J(t) \approx \frac{\Gamma_\mathrm{eff}(t)}{2}(\Delta \rho -\Delta \rho_\mathrm{th}(t)),
\label{Eq:Prob_Current2}
\end{equation}
where $\Gamma_\mathrm{eff}(t)$ is an instantaneous effective spin-flip rate, which we explicitly derive below.

\noindent The population difference $\Delta \rho$ is then found by solving the differential equation \mbox{$\Delta \dot \rho = -2J(t)$},
\begin{equation}
\begin{split}
	\Delta& \rho(t) = \exp \left[ -\int_{-\infty}^t \mathrm{d}t'\, \Gamma_\mathrm{eff}(t')\right] \bigg( \Delta \rho(-\infty)   \\
	&+ \int_{-\infty}^{t} \mathrm{d}t' \Delta\rho_\mathrm{th}(t') \Gamma_\mathrm{eff}(t') \exp \left[ + \int_{-\infty}^{t'} \mathrm{d}t'' \, \Gamma_\mathrm{eff}(t'') \right] \bigg).
\end{split}
\label{Eq:Diff_Eq_Delta_rho}
\end{equation}
This result essentially differs from that of Ref.~[\onlinecite{Leuenberger2000a}] by the second line, which allows for the electronic spins to equilibrate beyond equal population, \textit{i.e.} $\Delta \rho(t=\infty) <0$.

\noindent To evaluate $\Gamma_\mathrm{eff}(t)$, we first determine, for any given $\epsilon_B(t)$, the lowest sector $i_p(t)$ for which hybridization is relevant, \textit{i.e.}\ $\Delta_{i_p(t)} \gtrsim |\epsilon_B(t)|$. To flip the spin, the system should then climb to the configuration $A=(i_p-1,\uparrow)$. From there it transits via one of the two hybridized states $C_{\tau =\pm} = (i_p,\tau)$ and then descends via the state $B=(i_p-1,\downarrow)$. A simple, but lengthy calculation analogous to Ref.~\cite{Leuenberger2000} yields that at the selected time, the effective rate $\Gamma_{\rm eff}(t)$ evaluates to 
\begin{equation}
\begin{split}
	\Gamma_{\rm eff}(t)= \Gamma_{i_p}(t) &=
    \sum_{\tau = \pm}   \frac{e^{-\beta (\omega_{C_\tau}-\omega_{(0,\uparrow)})} + e^{-\beta (\omega_{C_\tau}-\omega_{(0,\downarrow)})}}{\gamma_{{C_\tau}A}^{-1}+\gamma_{{C_\tau}B}^{-1}}.  
\end{split}
\label{Eq:Gamma_Eff,i,Full}
\end{equation}
 
Note that the inverse of the denominator associated with passage through state $C_\tau$ is the sum of the expected times to transition from state $A=(i_p-1,\uparrow)$ to $C=(i_p,\tau)$ and then subsequently to state $B=(i_p-1,\downarrow)$.

\noindent At a given time $t$, the relaxation rate is typically dominated by a single pathway through the sector $i_p(t)$. While the time window $\Delta t_{i_p} \sim \Delta_{i_p}$, where a given $i_p(t)$ dominates, is much larger for larger $i_p$, its rate $\Gamma_{i_p}(t=0) \propto e^{-\beta \epsilon_{i_p}}$ (see below) is smaller. Generically, one of the total rates $\int \mathrm{d}t \, \Gamma_{i_p}(t)$ dominates the global spin-flip rate. Thus, we can approximate the total spin-flip rate as
\begin{equation}
	\Gamma_\mathrm{eff}(t) \approx \sum_{i_p=1}^{3} \Gamma_{i_p}(t).
\label{Eq:Gamma_Eff}
\end{equation}


\noindent \textbf{Spin-flip probability for small $\mathbf{B_x}$.} We now derive an approximate expression of $\Gamma_\mathrm{eff}$ in the limit of small applied fields $B_x$, where $\Delta_i \ll T$. Time-dependence in the transition rates enters only via the matrix elements between the instantaneous eigenstates, see Eq.~\eqref{Eq:Psi_H0}, 
\begin{equation}
\begin{split}
	\gamma_{(i_p-1,\uparrow/\downarrow), (i_p, \tau)} &\approx \gamma_{(i_p-1,\uparrow/\downarrow), (i_p, \uparrow)} |\braket{i_p,\uparrow|i_p,\tau}|^2  \\
	&\equiv \kappa \, p(I_{i_p}) m_i \left( 1 \pm \tau \frac{2 \epsilon_B(t)}{\sqrt{(2\epsilon_B(t))^2 + \Delta_{i_p}^2}} \right),
\end{split}
\label{Eq:Phonon_Rates_Approx}
\end{equation}
\begin{equation}
	m_i \equiv \left\lvert \frac{\braket{I_z=-7/2+(i-1)|I_-|I_z=-7/2+i}}{ \braket{I_z=-7/2|I_-|I_z=-5/2}} \right\lvert^2.
\label{Eq:Mat_El_Ratio}
\end{equation}

\noindent In the relevant tunneling window we can neglect the energy difference of the $i=0$ states , \mbox{$e^{-\beta (\omega_{(0,\uparrow)}-\omega_{(0,\downarrow)})} \approx 1$.} We then find the spin-flip rates
\begin{equation}
\begin{split}
	\Gamma_{i_p}(t) &
	\approx \frac{\Delta_{i_p}^2}{4 \epsilon_B^2(t) + \Delta_{i_p}^2}  (e^{-\beta \epsilon_{i_p}} \kappa \,m_{i_p}).
\label{Eq:Gamma_Eff_i}
\end{split}
\end{equation}
The Lorentzian factor \mbox{$\propto \Delta^2/((2\epsilon_B)^2 + \Delta^2)$} comes from the transition matrix elements to the entangled  instantaneous eigenstates, see Eq.~\eqref{Eq:Phonon_Rates_Approx}.  It is essentially the probability that the transition to the $i_p$ states  flips the electronic spin, while the subsequent decay does not involve a spin-flip, or vice versa.

\noindent Finally, using the spin-flip rate in the solution~\eqref{Eq:Diff_Eq_Delta_rho}, we can calculate the total spin-flip probability $P_\mathrm{flip}$ when tuning the external field through the zero-crossing. In the relevant Zeeman-energy window we can set $\Delta \rho_\mathrm{th}(t')\approx 0$, which yields~\footnote{The integration limit $t=+\infty$ is a valid approximation as an approximation extending the relevant integration window $0 \ll |\epsilon_B(t)| \lesssim \Delta$. For larger fields, the Hamiltonian~\eqref{Eq:H0} no longer applies since the relaxation dynamics is dominated by the HF crossing at $B_z \approx 230$\,Gs. Note that the divergence $\Gamma_\mathrm{eff}\propto B$ is irrelevant in the window $|\epsilon_B| \lesssim \Delta$, which justifies the use of the approximation $\Gamma_\mathrm{eff}$ in Eq.~\eqref{Eq:Gamma_Eff_i} and setting the integration limit to infinity.}
\begin{equation}
\begin{split}
	P_\mathrm{flip}\equiv& \rho_{(\downarrow,7/2)}(t=\infty) = \frac{1}{2} \left[1 -\Delta \rho(t=\infty) \right] \\
	\approx& \frac{1}{2} \left[1- \exp \left( -\int_{-\infty}^\infty \Gamma_\mathrm{eff}(t) \mathrm{d}t \right)\right] \\
	\approx& \frac{1}{2} \left[1- \exp \left(-\frac{\pi}{2} \sum_{i=1}^4 \frac{\Delta_i \, e^{-\beta \epsilon_i} \kappa m_i}{v} \right) \right].
\end{split}
\label{Eq:P_Flip}
\end{equation}

This result can be interpreted as a Landau-Zener probability,\cite{Zener1932} where the relevant tunneling time is $\Delta_i/v$ for each crossing $i$, with an attempt rate  $e^{-\beta \epsilon_i}\kappa m_i$. The spins can at most equilibrate to equal population, corresponding to a spin-flip probability $P_\mathrm{flip} = 1/2$, because the relevant energy window for thermalization is $|\epsilon_B|\sim \Delta_i\ll T$.

\noindent The spin-flip rate~\eqref{Eq:Gamma_Eff_i} and probability~\eqref{Eq:P_Flip} are in agreement with the theory of Refs.~[\,\onlinecite{Leuenberger2000},\onlinecite{Leuenberger2000a}] in the limit $\Delta_i \gg \kappa$, which applies to the dominant term  ($i=2$) in Eq.~\eqref{Eq:P_Flip} at the experimental temperatures. That path dominates because in the absence of an external transverse field (or at small applied fields), the larger gap compared to the $i=1$ states compensates for the additional activation cost, $\Delta_2/\Delta_1 \gg e^{-\beta E_\mathrm{HF}}$. While the gap of the $i=3$ states is yet a bit larger, it cannot mitigate the activation cost, $\Delta_3/\Delta_2 \ll e^{-\beta E_\mathrm{HF}}$. 

\noindent \textbf{Spin-flip probability for large $\mathbf{B_x}$.} For large transverse fields with $\Delta_i \gtrsim T$, the addition of the second term in the solution~\eqref{Eq:Diff_Eq_Delta_rho} 
becomes relevant. The effective rate $\Gamma_\mathrm{eff}$ is now non-negligible in a Zeeman-energy window $\epsilon_B \sim \Delta_i \gtrsim T$, which allows for rapid thermal equilibration at fields where the Zeeman splitting of the $i=0$ states is resolved and thus $\Delta \rho_\mathrm{th} \neq 0$. In this case, the spin-flip probability can exceed $1/2$, unlike in Eq.~\eqref{Eq:P_Flip}. At the final time $t_f$, we find the spin-flip probability
\begin{equation}
\begin{split}
	&P_\mathrm{flip}(t_f) \equiv \frac{1}{2} \left[1 -\Delta \rho(t_f) \right] =\\
	=&\frac{1}{2} \bigg[ 1 - \exp\left(- \int_{-\infty}^{t_f} \mathrm{d}t'\, \Gamma_{\mathrm{eff}}(t') \right) 
	- \int_{-\infty}^{t_f} \mathrm{d}t' \, \Delta \rho_\mathrm{th}(t') \Gamma_{\mathrm{eff}}(t') \exp\left( -\int_{t'}^{t_f} \mathrm{d}t''\, \Gamma_{\mathrm{eff}}(t'') \right)   \bigg], \\
\end{split}
\label{Eq:P_Flip_Exact}
\end{equation}

\noindent with $\Delta \rho_\mathrm{th}(t) = -\tanh(\beta \epsilon_B(t))$. 

\noindent The result~\eqref{Eq:P_Flip_Exact}, including the spin-flip rate $\Gamma_\mathrm{eff}(t)$, has to be evaluated numerically for $\Delta_i \gtrsim T$. Nevertheless, it can be easily seen that the spin-flip probability can indeed exceed $P_\mathrm{flip}>1/2$ by considering the limit $\Delta_{i} \gg T$. Since in most of the relaxation window one finds $\beta |\epsilon_B| \gg 1$, we can approximate $\Delta \rho_\mathrm{th}(t) = -\text{sgn}(t)$. Then, Eq.~\eqref{Eq:P_Flip_Exact} reduces to 
\begin{equation}
\label{Eq:P_Flip_Approx}
	P_\mathrm{flip}(t_f) \approx 1- \exp\left(\int_0^{t_f} \mathrm{d}t \, \Gamma_\mathrm{eff}(t) \right).
\end{equation}
At long times  $t_f\gg k_B T/(2v)$ -- where the Zeeman splitting is larger than the temperature --  and for small sweep rates $v \ll \max_{i>0}\left[ \Delta_{i} e^{-\beta \epsilon_{i}}\kappa m_{i}\right]$ (see Eq.~\ref{Eq:Gamma_Eff_i}), the spins have flipped with probability $P_\mathrm{flip} \approx 1$. In contrast to the expressions in Refs.~[\,\onlinecite{Leuenberger2000}, \onlinecite{Leuenberger2000a}], the additional correction term in Eqs.~(\ref{Eq:Diff_Eq_Delta_rho},\ref{Eq:P_Flip_Exact}) keeps the system in thermal equilibrium for $v \to 0$, as one expects for adiabatic driving.

\subsubsection{$\mathbf{B_z \approx \pm 230}$\,Gs}

A theory of activated tunneling can analogously be worked out for the crossing at 230\,Gs. The conclusion is that the avoided gap in the sector $i_p=2$ is so large, and thus activated tunneling so efficient, that at all experimental sweep rates the system should equilibrate internally. However, the phonons may not relax sufficiently rapidly to the external bath temperature due to a phonon bottleneck. 

\subsection{The phonon bottleneck}

During the field sweep the phonons may often fall out of equilibrium with the external heat bath because the number of relevant phonon modes with energies in the range of nuclear spin transitions is several orders of magnitude smaller than the number of Ho ions. Their ratio $N_\mathrm{ph}/N_\mathrm{Ho}$ can be estimated as~\cite{Barbara:1999fk}
\begin{equation}
	\frac{N_\mathrm{ph}}{N_\mathrm{Ho}} = \frac{\rho(\omega) \Delta \omega}{4x/V_\mathrm{uc}} = 3.0 \times 10^{-7},
	\label{Eq:Bottleneck}
\end{equation}
where $\rho(\omega) =3  \omega^2/(2\pi^2c^3)$ is the Debye density of states of the phonons at the relevant transitions frequency of $\hbar \omega/k_B = 0.2 \text{\,K}$, $x=0.3\%$ is the Ho concentration and $V_\mathrm{uc}$ is the unit cell volume of LiHoF$_4$ containing four Ho ions. We use the phonon sound velocity in LiYF$_4$ of \mbox{$c= 3 \times 10^3 \text{\,m/s}$}~\cite{Blanchfield1979} and the linewidth of HF levels of $\Delta \omega =  3.2 \text{\,GHz}$\cite{Beckert:2020pi}.
\end{methods}

\bibliography{Spin_tunneling_LiHoF4}

\begin{addendum}
\item[Competing Interests] The authors declare that they have no competing financial interests.
\item[Correspondence] Correspondence and requests for materials should be addressed to G.~M.\\ (email: guy.matmon@psi.ch).
\end{addendum}

\appendix

\section{Ground-state level spectrum in a longitudinal field}	 

\begin{figure}[H]
	\centering
	\begin{tabular}{p{0.34\textwidth}p{0.28\textwidth}p{0.275\textwidth}}
		\subfloat{
            \hspace{-11mm}
				\stackinset{l}{45pt}{t}{-10pt}{(a)}{\includegraphics[scale =0.9] {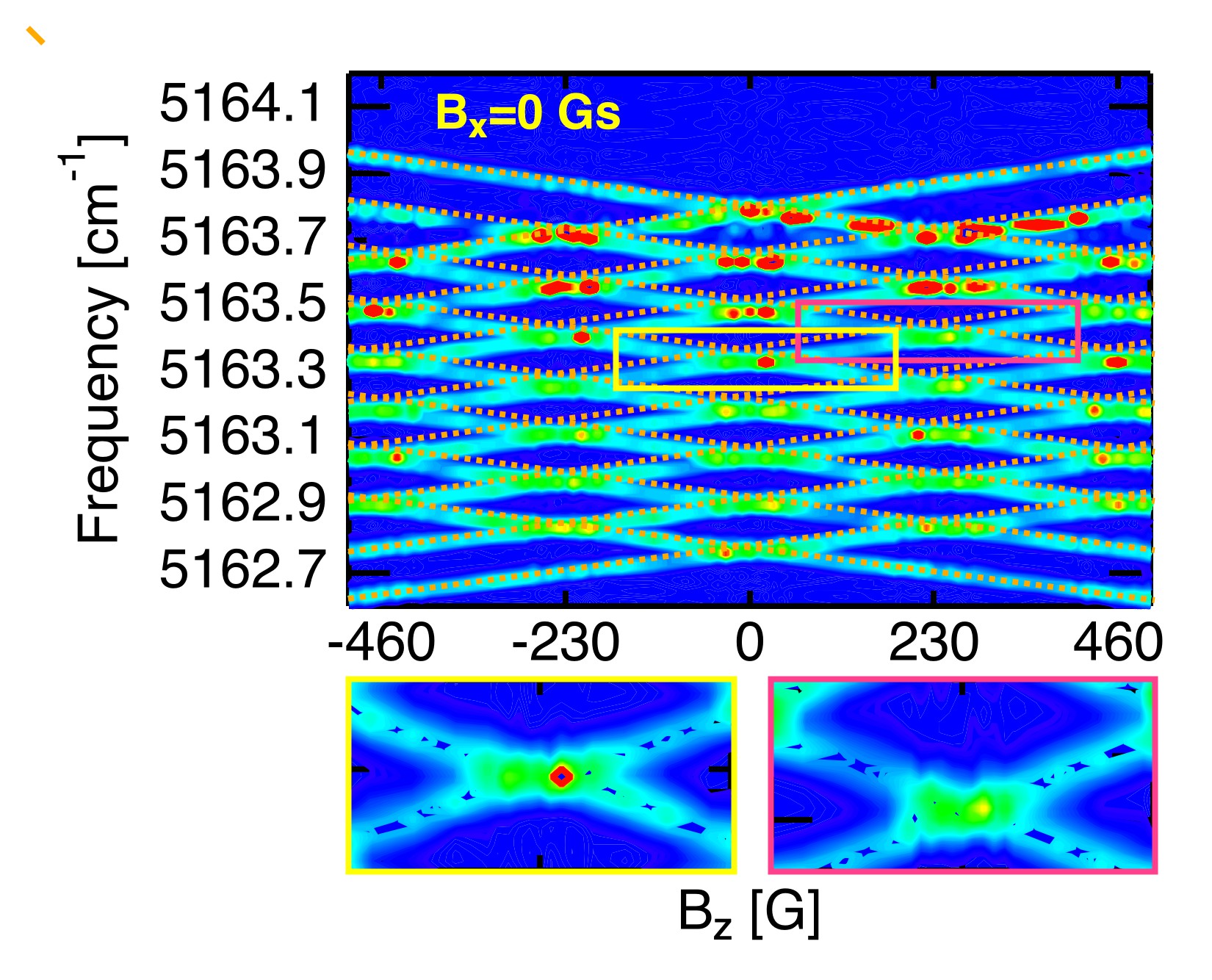}}
			\label{Fig:Long_Field_0_Trans_Inset}
		}&
		\subfloat{
            \hspace{-7mm}
			\stackinset{l}{0pt}{t}{-10pt}{(b)}{\includegraphics[scale =0.9] {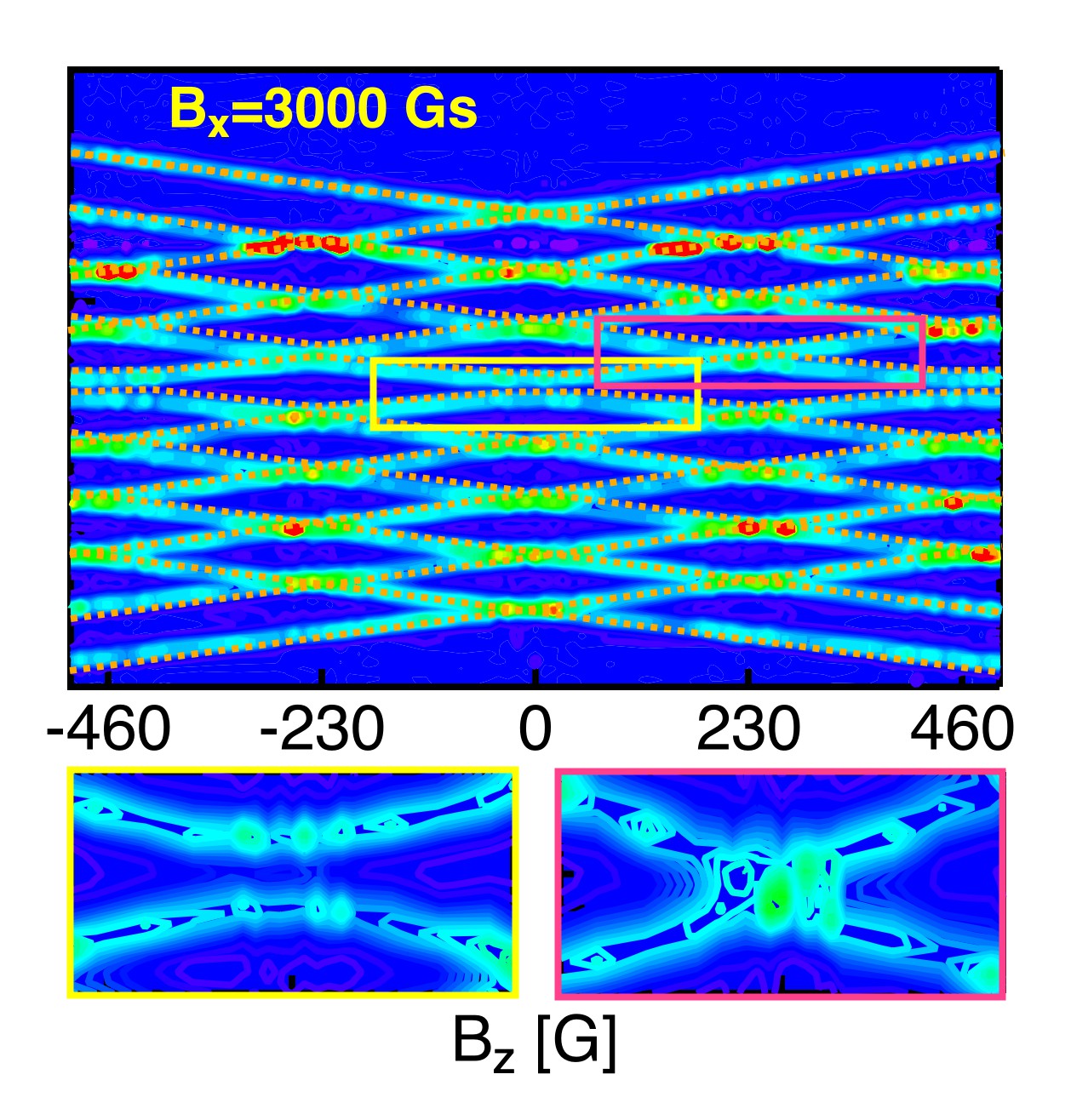}}
				\label{Fig:Long_Field_3000_Trans_Inset}
		}&
		\subfloat{
            \hspace{-9mm}
			\stackinset{l}{0pt}{t}{-10pt}{(c)}{\includegraphics[scale =0.9] {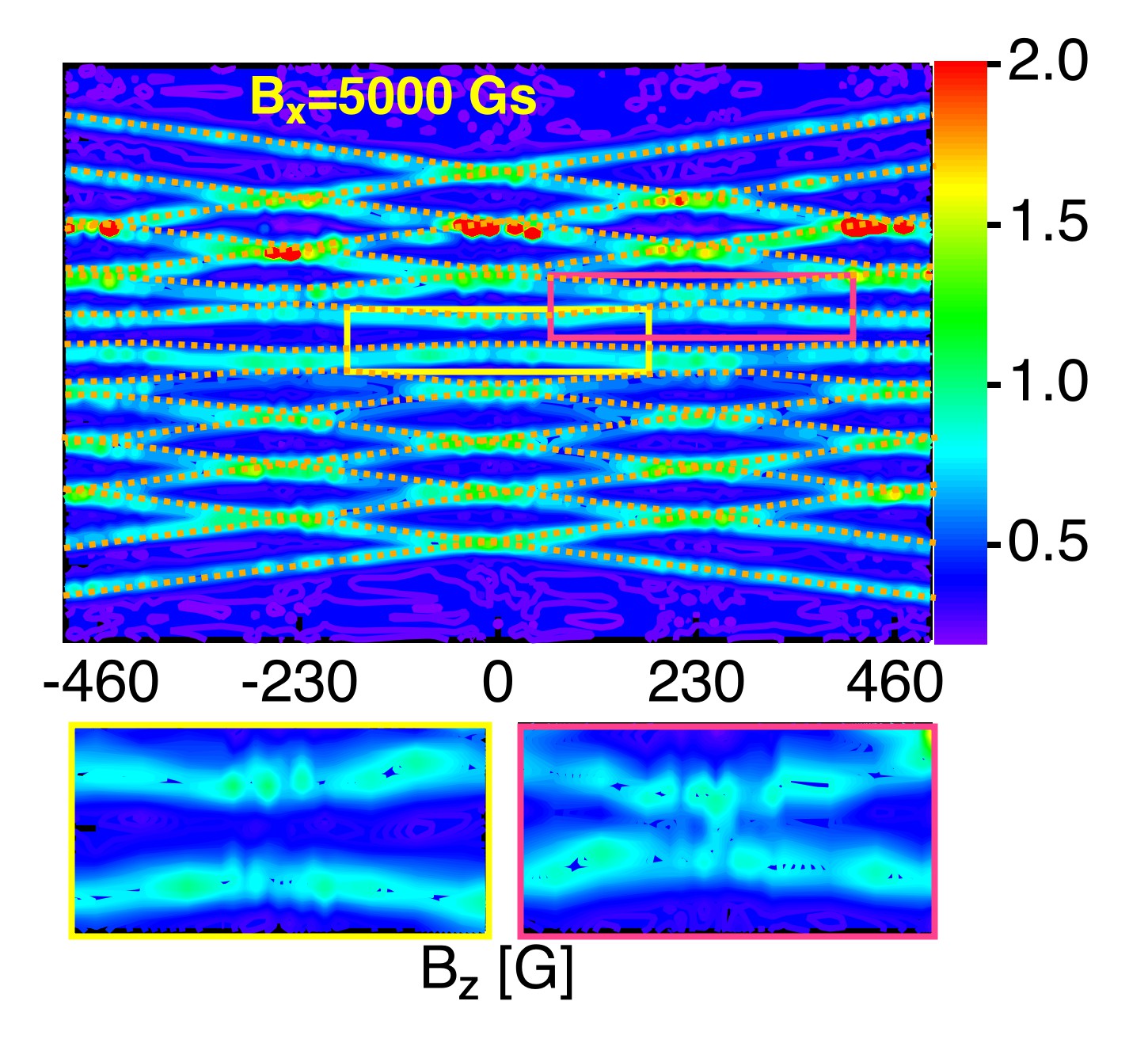}}
			\label{Fig:Long_Field_5000_Trans_Inset}
		}
	\end{tabular}
	\caption{HF levels in a longitudinal magnetic field sweep. \protect\subref{Fig:Long_Field_0_Trans_Inset} -- \protect\subref{Fig:Long_Field_5000_Trans_Inset} Longitudinal field sweep with different transverse fields. For each transverse field two crossings are blown up, one with $\Delta I_z=-1$ (yellow frame) and one with $\Delta I_z=-2$ (magenta frame). All sweeps were done at 10\,K. The color axis units are absorbance. The dotted yellow lines are the calculated HF levels. Their global spectrum shifts by an amount $\propto B_x^2$ due to the ground state doublet being level-repelled from close CF singlets (see section \ref{sec:second_order} in the SI).}
	\label{Fig:Long_Field_Inset}
\end{figure}

Fig.~\ref{Fig:Long_Field_Inset} shows the measured and calculated absorption lines as a function of longitudinal field. The small deviations between the numerical and experimental results are due to neglect of quadrupolar HF corrections to the doublet states, and of HF corrections to the final singlet state. Note that the highest transition energy corresponds to the excitation from the lowest energy doublet state. As the final state in the $^5I_7$ manifold we use the singlet at energy $5163\,{\rm cm}^{-1}$, as it acquires only a minimal moment in response to a transverse field, and thus does not introduce a noticeable additional HF splitting.

\section{Ground-state level spectrum in a transverse field}	 

\begin{figure}[ht!]
	\centering
	\begin{tabular}{p{0.345\textwidth}p{0.285\textwidth}p{0.28\textwidth}}
	    \subfloat{
		\hspace{0mm}
			\stackinset{l}{0pt}{t}{-10pt}{(a)}{\includegraphics[width =1.4\linewidth,valign=b] {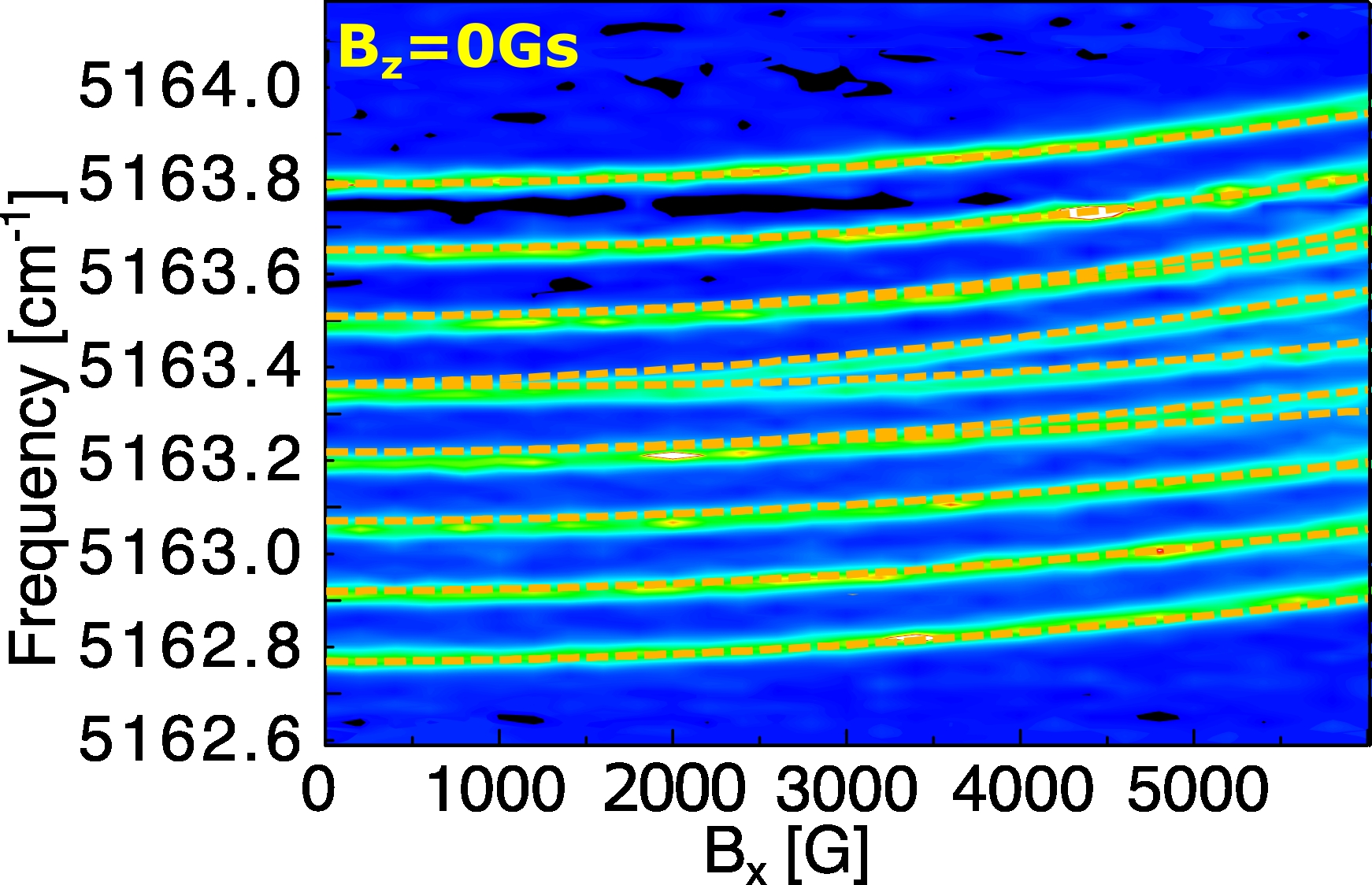}}
			\label{Fig:Trans_Field_0_Long_SI}
		}&
        &
        \subfloat{
		\hspace{-30mm}
			\stackinset{l}{0pt}{t}{-10pt}{(b)}{\includegraphics[width =1.54\linewidth,valign=b] {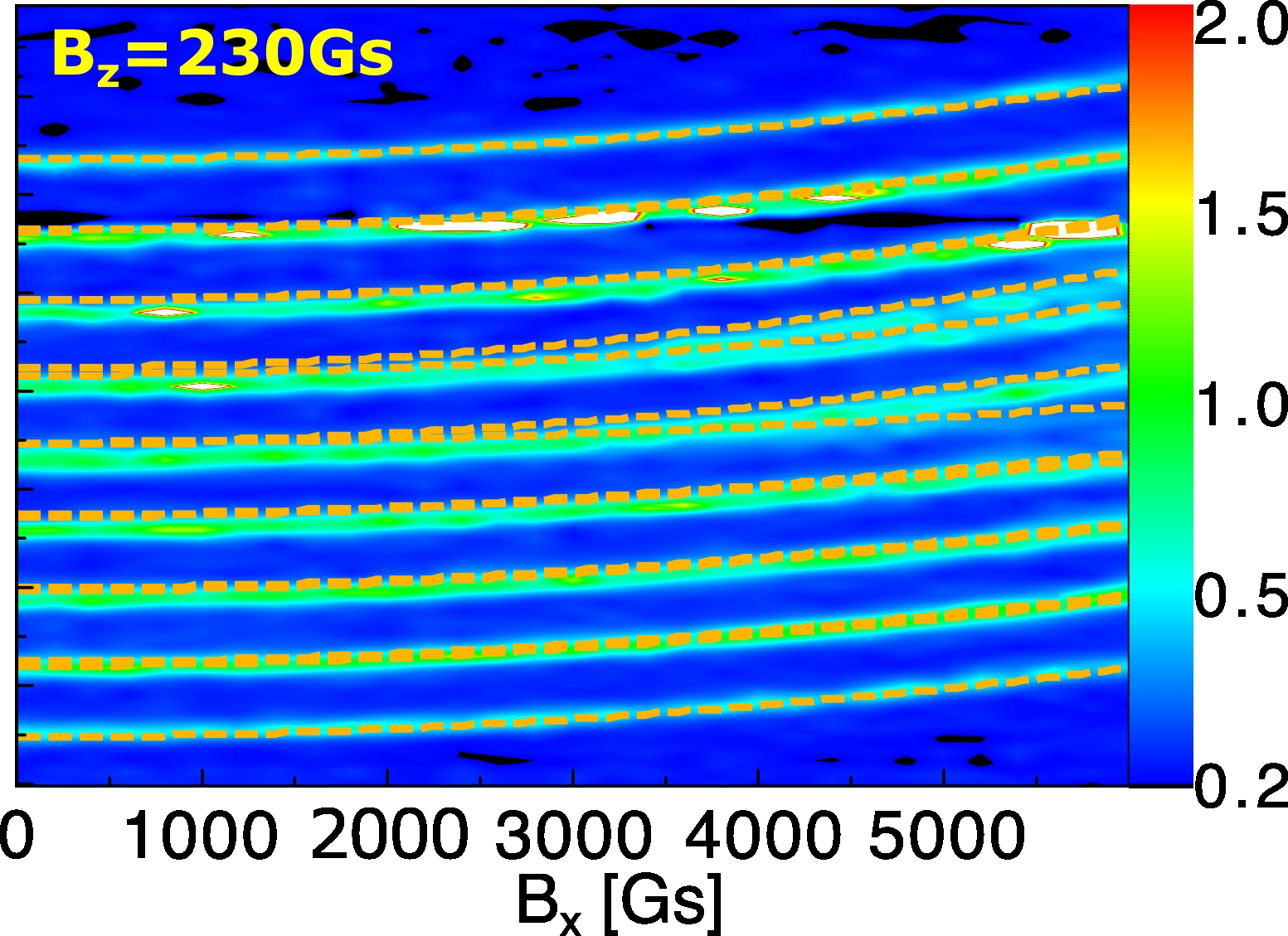}}
			\label{Fig:Trans_Field_230_Long_SI}
		}
	\end{tabular}
	\caption{HF levels in a transverse magnetic field sweep. \protect\subref{Fig:Trans_Field_0_Long_SI} and \protect\subref{Fig:Trans_Field_230_Long_SI}: transverse field sweep with different longitudinal fields.}
	\label{Fig:Trans_Field_SI}
\end{figure}

Fig \ref{Fig:Trans_Field_SI} shows the measured and calculated absorption lines as a function of transverse field.

\section{Data on spectroscopic maps of the hysteresis}

\begin{figure}
    \centering
    \includegraphics[width=1\linewidth]{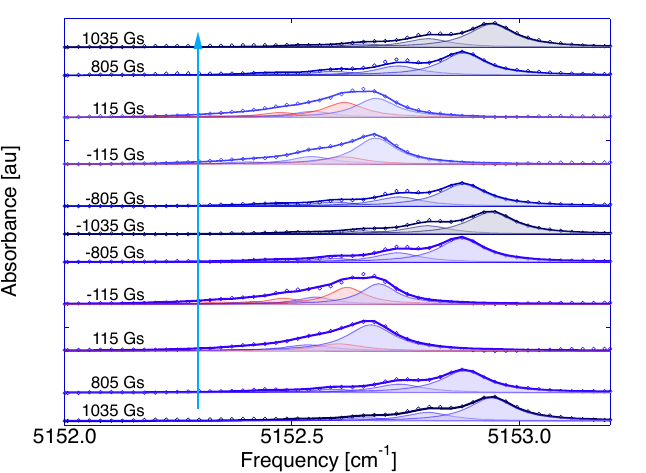}
    \caption{The spectra from the hysteresis curve in Fig. \ref{Fig:Rotation_Diff_Bx}. The fits are constrained by the magnetizations in Table \ref{tab:T_M}, which were measured independently. The equivalence between spectra before the zero crossings (the lower 115\,Gs and the higher -115\,Gs) and after (the lower -115\,Gs and the higher 115\,Gs) can be seen.}
    \label{Fig:50mK_Fits_Full}
\end{figure}

All ten measured hysteresis spectra and their constituent HF lines appear in Fig. \ref{Fig:50mK_Fits_Full}.

A summary of magnetization and $T_n$ for each measurement point along the hysteresis curves (blue markers in the inset of Fig. \ref{Fig:Faraday_Magnetization}) is given in table \ref{tab:T_M}. 

\begin{table}[H]
    \footnotesize
    \centering
    \renewcommand{\arraystretch}{0.6}
    \begin{tabular}{ccc||ccc}
          $\text{B}_\text{z}$\,[Gs] & Magnetization & $T_n$\,[mK] & $\text{B}_\text{z}$\,[Gs] & Magnetization & $T_n$\,[mK]\\
          \hline
           1035& 1.000(48) & 202(20) & -1035& -1.000(46) & 211(15)\\
           805& 1.000(42) & 218(11) & -805& -1.000(46) & 231(18)\\
           115& 0.729(44) & 138(10) & -115& -0.719(40) & 163(7)\\
           -115& -0.174(43) & 192(18) & 115& 0.214(44) & 174(12)\\
           -805& -1.000(48) & 217(20) & 805& 0.985(46) & 237(22)\\
    \end{tabular}
    \caption{Measured magnetization and fitted effective nuclear spin temperatures $T_n$ at different longitudinal fields $B_z$ along the hysteresis loop (Figs. \ref{Fig:Faraday_Magnetization} and \ref{Fig:50mK_Fits}).}
    \label{tab:T_M}
\end{table}

\section{Sample temperature during hysteresis loops}
\label{Sec:SI_Temperature}

\begin{figure}[b!]
	\centering
	\includegraphics[width = 0.8\linewidth] {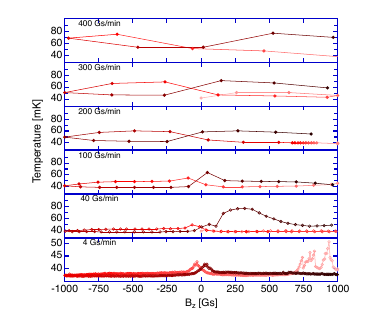}
	\caption{Temperature measurement as a function of $B_z$ during the hysteresis sweep (back and forth, without pause at the turning point) for different sweep rates. For each sweep rate time progresses from bright (start) to dark (end) color. The temperature sensor is sampled at fixed time intervals due to system limitations, therefore faster sweeps have fewer points.}
	\label{Fig:T_vs_B}
\end{figure}

\begin{figure}[b!]
	\centering
	\includegraphics[width = 0.8\linewidth] {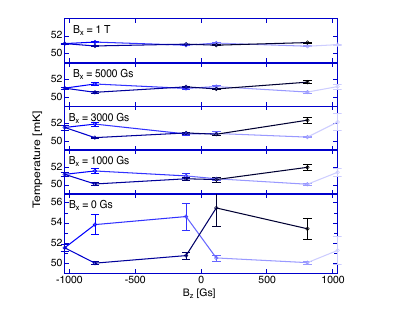}
	\caption{Temperature measurement as a function of $B_z$ during the FTIR hysteresis sweep for different transverse fields at constant field sweep rate. For each sweep rate time progresses from bright (start) to dark (end) color. As the transverse field increases the temperature hysteresis closes.}
	\label{Fig:T_vs_Bz_diff_Bx}
\end{figure}

We measured the temperature of the sample during the sweeps with a sensor attached to the sample (Figures \ref{Fig:T_vs_B} and \ref{Fig:T_vs_Bz_diff_Bx}). 
As  expected, apart from an initial transient, forward and backward sweeps are  symmetric. The thermometer shows an elevated phonon temperature that varies along the hysteresis loop. 
In the absence of illumination, this arises from two sources: (1) sweep-induced eddy currents constantly heat the system (the total injected energy being proportional to the sweep rate) and 
(2) The temperature starts to rise before the large drop at $B_z=0$ is reached, at a field that seems largely independent of sweep rate. This must be due to dissipation that arises because of spin-spin interactions: Certain spins in negative internal fields flip before $B_z=0$ (the occurrence of such spins is just a function of concentration, not of sweep rate).
Their flipping can destabilize other spins, putting them into an unstable configuration from which a flip of an electron spin involving a phonon emission can occur. The faster the sweep rate, the larger the external field tends to be when this dissipative flip occurs, and thus the stronger is the heating. This explains the temperature bump that starts before the $B_z=0$ crossing and becomes maximal shortly afterwards, where such flip events become frequent. Note  that in case significant energy is dissipated due to magnetic hysteresis, there is no guarantee that the phonons inside the sample are equilibrated with the sensor. Indeed, we have reasons to believe that under illumination and in fast sweeps at low $T$ the phonons are hotter than the sensor reading suggests. 

\noindent Beyond $-230G$, the temperature relaxes to a steady value set by current heating and continuous spin relaxation. The latter diminishes at larger field and $T$ decreases.

\noindent The application of transverse field (Figure \ref{Fig:T_vs_Bz_diff_Bx}) modifies the dynamics above. As the electron-phonon coupling is enhanced due to the stronger admixtures of the singlet CF state and the stronger phonon matrix element between the latter and the ground doublet, the electron spins come closer to thermal equilibrium at all values of $B_z$.

\noindent Conversely, when HF populations were measured via Faraday rotation with FTIR light, the sample was significantly more excited. Indeed the hysteresis curve in light was similar to one in the dark but at $T$ stabilized to 120mK ($=T_e$), rather than 40mK. Since electron flips away from level crossings involve phonons, it is likely that this value also reflects a similarly elevated temperature of the phonons within the sample which thus exceeds the temperature measured outside. The fit of the HF spectra to an effective nuclear spin temperature, instead yielded $T_n \approx 200\,\text{mK}$. 

\section{Modeling of hysteresis}

\subsection{Thermalization from hysteresis measurements.}
\label{Sec:Hyst_Therm}

\begin{figure}[H]
    \centering
	\subfloat[]{
        \includegraphics[width=0.51\linewidth,valign=t]{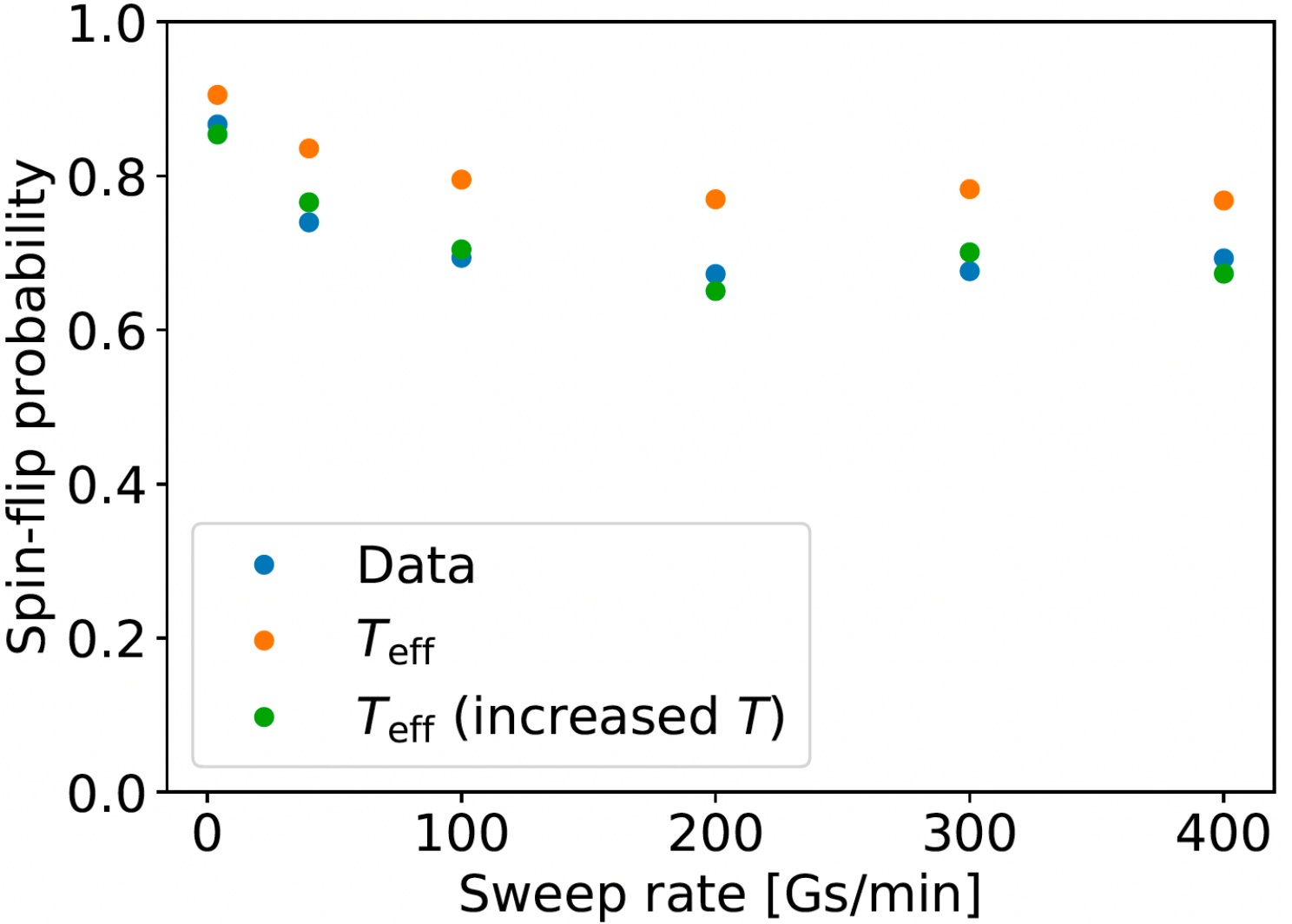}
        \label{Fig:Mag_Drop_230Gs_Sweep_Rate}
    }
	\subfloat[]{
        \includegraphics[width=0.45\linewidth,valign=t]{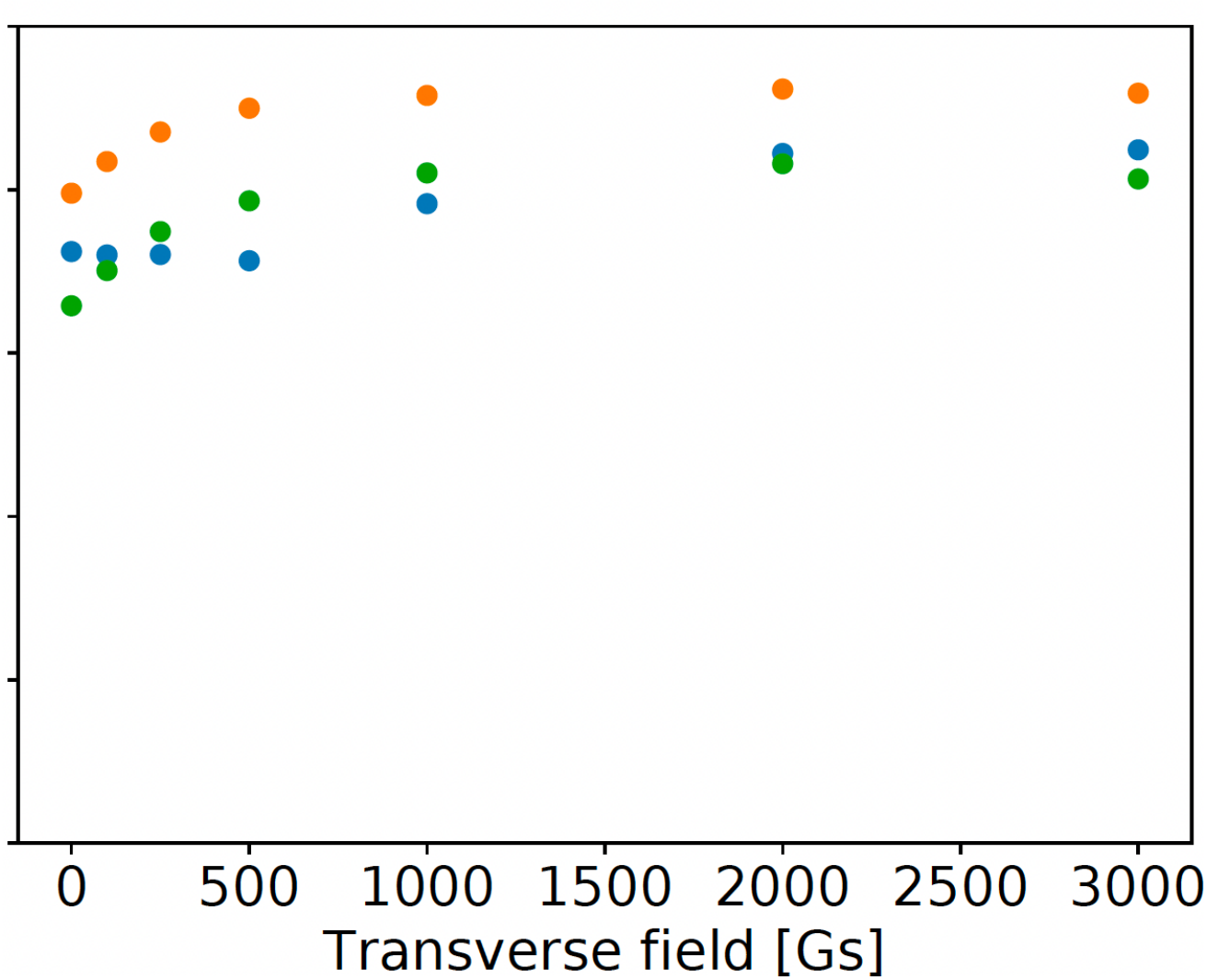}
        \label{Fig:Mag_Drop_230Gs_Bx}
    }
    \caption{Magnetization (blue) after crossing $B_z = 230\,\text{Gs}$ for the sweep from negative to positive field as a function of \protect\subref{Fig:Mag_Drop_230Gs_Sweep_Rate} sweep rate and \protect\subref{Fig:Mag_Drop_230Gs_Bx} transverse field. We compare the data to the magnetization expected from the effective temperature $T_\text{eff}$ (orange) as defined via the energy in the spin system before the magnetization drop. For the green dots, we have increased the nuclear-spin temperature as compared to the bath temperature by a factor \protect\subref{Fig:Mag_Drop_230Gs_Sweep_Rate} 2.7 and \protect\subref{Fig:Mag_Drop_230Gs_Bx} 3.2 (best fit values).}
    \label{Fig:Mag_Drop_230Gs}
\end{figure}

Analysis of Fig. \ref{Fig:Hyst} can qualitatively inform us about the thermalization of the different populations (or lack thereof) even without detailed spectral data. In Fig. \ref{Fig:Mag_Drop_230Gs} 
we compare the measured magnetization drop at $B_z = 230\,\text{Gs}$ (extracted at $B_z = 300\,\text{Gs}$) to the one expected if the energy in the spin system is conserved and distributes ergodically over all spin degrees of freedom. 
Let us  provide evidence that the nuclear spins are hotter than the bath when the magnetization drop at $B_z = 230\,\text{Gs}$ is approached.  
We show this by demonstrating that the converse assumption, namely that the nuclear spins are equilibrated before the crossing (with the bath’s temperature), is inconsistent with the data. We find that the drop in the magnetization is consistently smaller than expected from the effective spin temperature $T_e=T_n= T_\text{eff}$ one expects to establish. This suggests that $T_\text{eff}$ is in fact larger and that there is therefore extra energy in the system that we did not account for. This remaining discrepancy is likely  explained by a larger temperature of the hyperfine states as compared to the measured temperature, which increases the energy of the spin system and thus $T_\text{eff}$. To obtain a good agreement with the experimentally observed drop in magnetization we need to assume an effective temperature of the hyperfine states (right before $B_z = 230\,\text{Gs}$) that exceeds the measured phonon temperature by a factor of $\approx 3$.

The origin of such an elevated nuclear-spin temperature is presumably a combination of the heating occurring around $B_z = 0$, the continuous heating due to delayed electron spin relaxation away from level crossings, and a hyperfine population that is already out of equilibrium at the start of the experiment, \textit{i.e.} at very negative (or positive) $B_z$. The spectroscopic measurements whose results are summarized in Table \ref{tab:T_M} suggest that the latter is the case

\section{Perturbation theory for avoided level crossings}
\label{Sec:Perturbation_Theory}

\subsection{Gaps of avoided level crossings}
Here we derive analytical expressions for the gaps of HF crossings in the ground-state doublet. We use degenerate perturbation theory, treating the transverse field and the transverse HF interaction in the Hamiltonian~(\ref{Eq:H_Single_Ion}) as perturbations. To leading order in perturbation theory, the longitudinal field and the longitudinal HF interaction need to be taken into account only to first order through the shift of the energy levels. The unperturbed states are denoted as $\ket{\Gamma^{(i)}, I_z} \equiv \ket{\Gamma^{(i)}} \otimes \ket{I_z}$. The CF states in the lowest $J$-manifold that carry the same representation $\Gamma$  are numbered by a superscript $i\geq 0$ that increases with the energy.
The associated CF energy is denoted as $\varepsilon_{\Gamma}^{(i)}$ (as measured from the ground-state doublet, $\epsilon_{\Gamma_{3,4}}^{(0)} \equiv 0$). For the ground-state doublet we adopt the notation from above, $\ket{\uparrow} \equiv \ket{\Gamma_3^{(0)}}$ and $ \ket{\downarrow} \equiv \ket{\Gamma_4^{(0)}}$. 

\noindent We choose the complex  phases of the electronic CF states such that under time reversal, $\mathcal{T}$, the one-dimensional representations are invariant, \mbox{$\mathcal{T} \ket{\Gamma_{1/2}^{(i)}} = \ket{\Gamma_{1/2}^{(i)}}$}, while the states of a doublet representation are exchanged, \mbox{$\mathcal{T} \ket{\Gamma_{3/4}^{(i)}} = \ket{\Gamma_{4/3}^{(i)}}$}. This choice leads to simple relations between matrix elements involving states $\ket{\downarrow}$ and $\ket{\uparrow}$ that are related by time-reversal.

\noindent Using this notation, we derive below the avoided gaps $\Delta \varepsilon(B_n, \Delta I_z)$ associated with the crossing fields $B_{n}$ in Eq.~\eqref{Eq:Bz_Crossing}, with \mbox{$\Delta I_z \equiv I_z^\downarrow- I_z^\uparrow $}. We consider only the leading order in (transverse) HF coupling $A$ and (subsequently)  leading order in transverse field $B_x$ through their higher-order admixing of excited CF levels.
In doing so, we neglect the HF splitting, \mbox{$E_\mathrm{HF}\!\approx\!\!A\,m_z\!\!\approx\!0.21\,\mathrm{K}$}, in those denominators that involve energy differences between different CF levels. We further exploit a particularity of \LiHoF: one  finds that the CF levels closest to the ground state are $\Gamma_2$ singlets, while the first $\Gamma_1$ singlet appears at much higher energy, implying the inequalities  \mbox{$\varepsilon_2^{(0)},\varepsilon_2^{(1)} \ll \varepsilon_1^{(0)},\varepsilon_1^{(1)}$}. We thus seek the leading perturbative expression involving as few  virtual $\Gamma_1$ states as possible and only retain that expression.

\subsection{Avoided crossings at $\mathbf{B_z=B_0=0}$}
At \mbox{$B_z=B_0=0$}, to lowest order in $A$ and $B_x$, and with the above approximations, we find the gaps of the four lowest electronuclear doublets  as
\begin{equation}
    \begin{aligned}
        	\Delta \varepsilon(B_0 = 0, \Delta I_z) \approx & 16\mu_B g_L B_x  \frac{A^{(|\Delta I_z|+1)/2}}{m_z^{(|\Delta I_z|-1)/2}}   
	\times\\
	\times&\begin{cases}
		\Big\lvert \sum_{i\geq0} \frac{\braket{\uparrow|J_x|\Gamma_2^{(i)}}^2}{ \varepsilon_2^{(i)}} \Big\lvert,& \text{for } \Delta I_z = 1,\\
		15 \Big\lvert  \sum_{i,j\geq0} \frac{\braket{\uparrow|J_x|\Gamma_2^{(i)}}^2 |\braket{\uparrow|J_x|\Gamma_2^{(j)}}|^2}{\varepsilon_2^{(i)} \varepsilon_2^{(j)}} \Big\lvert,& \text{for } \Delta I_z=3,\\
		50 \Big\lvert \sum_{i,j,k\geq0} \frac{\braket{\uparrow|J_x|\Gamma_2^{(i)}}^2 \braket{\uparrow|J_x|\Gamma_2^{(j)}}^2 \braket{\Gamma_1^{(k)}|J_x|\uparrow}^2}{\varepsilon_2^{(i)} \varepsilon_2^{(j)} \varepsilon_1^{(k)}} \Big\lvert,& \text{for } \Delta I_z=5,\\
		\frac{378}{5} \Big\lvert \sum_{i,j,k,l\geq0} \frac{\braket{\uparrow|J_x|\Gamma_2^{(i)}}^2 |\braket{\uparrow|J_x|\Gamma_2^{(j)}}|^2 \braket{\uparrow|J_x|\Gamma_2^{(k)}}^2 \braket{\Gamma_1^{(l)}|J_x|\uparrow}^2}{\varepsilon_2^{(i)} \varepsilon_2^{(j)} \varepsilon_2^{(k)} \varepsilon_1^{(l)}} \Big\lvert,& \text{for } \Delta I_z=7,
	\end{cases}
    \end{aligned}
    \label{Eq:Gap_Zero_Field_1}
\end{equation}
and the gaps of the four higher-lying electronuclear crossings (note the different intermediate singlet states)
\begin{equation}
    \begin{aligned}
    \Delta \varepsilon(B_0 = 0, \Delta I_z) \approx & 16\mu_B g_L B_x \frac{A^{(|\Delta I_z|+1)/2}}{m_z^{(|\Delta I_z|-1)/2} }
	\times\\
	\times & \begin{cases}
		\Big\lvert  \sum_{i\geq0} \frac{\braket{\uparrow|J_x|\Gamma_1^{(i)}}^2}{\varepsilon_1^{(i)}}  \Big\lvert,& \text{for } \Delta I_z = -1,\\
		15 \Big\lvert \sum_{i,j\geq0}   \frac{\braket{\uparrow|J_x|\Gamma_1^{(i)}}^2 |\braket{\uparrow|J_x|\Gamma_2^{(j)}}|^2}{\varepsilon_1^{(i)} \varepsilon_2^{(j)}} \Big\lvert,& \text{for } \Delta I_z=-3,\\
		50 \Big\lvert \sum_{i,j,k\geq0} \frac{\braket{\uparrow|J_x|\Gamma_1^{(i)}}^2 \braket{\Gamma_2^{(j)}|J_x|\uparrow}^2 \braket{\uparrow|J_x|\Gamma_1^{(k)}}^2}{\varepsilon_1^{(i)} \varepsilon_2^{(j)} \varepsilon_1^{(k)}} \Big\lvert,& \text{for } \Delta I_z=- 5,\\
		\frac{378}{5} \Big\lvert \sum_{i,j,k,l\geq0}  \frac{\braket{\uparrow|J_x|\Gamma_1^{(i)}}^2 |\braket{\uparrow|J_x|\Gamma_2^{(j)}}|^2 \braket{\Gamma_2^{(k)}|J_x|\uparrow}^2 \braket{\uparrow|J_x|\Gamma_1^{(l)}}^2}{\varepsilon_1^{(i)} \varepsilon_2^{(j)} \varepsilon_1^{(k)} \varepsilon_1^{(l)}} \Big\lvert,& \text{for } \Delta I_z= -7.
	\end{cases}
	    \end{aligned}
    \label{Eq:Gap_Zero_Field_2}
\end{equation}
The splitting is always linear in $B_x$, and thus vanishes for $B_x=0$, reflecting Kramers degeneracy. Note also the powers of $A$. The  exponent is a result of applying the transverse HF interaction $|\Delta I_z|$ times to flip the nuclear spin, which is diminished by the number $(|\Delta I_z|-1)/2$ of intermediate energy denominators that scale as $\propto A \, m_z$. Those arise from virtual HF states of the electronic ground-state doublet with intermediate values of $I_z$.

\subsection{Avoided crossings at finite $\mathbf{B_z\approx B_{\pm 1}=\pm 230}$\,Gs}

For crossings at the finite longitudinal field \mbox{$B_{\pm 1} = \pm 230\text{\,G}$}, we find for the gaps between states with $\Delta I_z  = 2 \,\mod\, 4$ in the absence of a transverse field 
\begin{equation}
\begin{aligned}
	\Delta \varepsilon (B_{\pm 1} \approx \pm 230 \text{\,G}, \Delta I_z) \approx & 8 \sqrt{15} \frac{A^{(|\Delta I_z|+1)/2}}{m_z^{(|\Delta I_z|-1)/2}} \times\\
	\times & \begin{cases}
		\Big \lvert \sum_{i\geq0} \frac{\braket{\uparrow|J_x|\Gamma_2^{(i)}}^2 }{\varepsilon_2^{(i)}} \Big\lvert,& \text{for } \Delta I_z= \pm 2, \\
		\Big \lvert \sum_{i\geq0}  \frac{ \braket{\uparrow|J_x|\Gamma_1^{(i)}}^2}{\varepsilon_1^{(i)}} \Big \lvert,& \text{for } \Delta I_z= \mp 2, \\
		 \frac{3\sqrt{105}}{4} \Big \lvert \sum_{i,j,k\geq0} \frac{\braket{\Gamma_1^{(i)}|J_x|\uparrow}^2 \braket{\uparrow|J_x|\Gamma_2^{(j)}}^2 \braket{\uparrow|J_x|\Gamma_2^{(k)}}^2 }{\varepsilon_1^{(i)} \varepsilon_2^{(j)} \varepsilon_2^{(k)}} \Big\lvert,& \text{for } \Delta I_z= \pm 6, \\
		\frac{3\sqrt{105}}{4} \Big \lvert \sum_{i,j,k\geq0} \frac{\braket{\Gamma_1^{(i)}|J_x|\uparrow}^2 \braket{\uparrow|J_x|\Gamma_2^{(j)}}^2 \braket{\Gamma_1^{(k)}|J_x|\uparrow}^2}{\varepsilon_1^{(i)} \varepsilon_2^{(j)} \varepsilon_1^{(k)}} \Big\lvert,& \text{for } \Delta I_z= \mp 6.
	\end{cases} 
\end{aligned}
\label{Eq:Delta_Iz_2,6}
\end{equation}
The leading exponent of $A$ is again a result of the number $|\Delta I_z|$ of transverse HF interactions and the $(|\Delta I_z|- 1)/2$ intermediate states whose energies scale as $\propto A m_z$.

\subsection{Second order energy shift in transverse field $\mathbf{B_x}$}
\label{sec:second_order}
Apart from lifting degeneracies between HF states, a transverse field $B_x$ also shifts the CF levels by admixing other CF states. In particular, the ground-state doublet $\ket{\uparrow},\ket{\downarrow}$ experiences a repulsion $\Delta \varepsilon(B_x)$ from the higher-lying singlets to second order in $B_x$,
\begin{equation}
	\Delta \varepsilon (B_x)\approx -(\mu_B g_L B_x)^2 \left\lvert \sum_{i\geq 0} \frac{\braket{\uparrow|J_x|\Gamma_2^{(i)}}^2 }{\varepsilon_2^{(i)}} \right\lvert.
\label{Eq:Second_Order_Shift}
\end{equation}
This downshift manifests itself in the spectrum in Fig.~\ref{Fig:Trans_Field_SI}, where the transition energies increase with the transverse field.

\section{Thermally activated quantum tunneling}

At any instantaneous field $B_z(t)$, we can solve Eqs.~\eqref{Eq:Rate_Eq_2} and \eqref{Eq:Prob_Current} to obtain the current $J(t)$. We do this explicitly below for our system, but first we analyze the more generic solution. Since the master equation is linear in  $\{\rho_{i,\sigma}\}$, the stationary limit ($\dot \rho_{i,\tau}=0$) of the excited states $i\geq 1$ implies that, by elimination, they can be expressed in terms of linear functions of $\rho_{(0,\uparrow)}$ and $\rho_{(0,\downarrow)}$ only. Inserting these expressions into the stationary flow equations for $i=0$, i.e.\ $\dot \rho_{0,\uparrow/\downarrow} = \pm J(t)$, yields that $J(t)$ can also be expressed as linear functions of $\rho_{(0,\uparrow)},\rho_{(0,\downarrow)}$. Since only the lowest $i=0$ states are considerably populated at the experimental temperatures, we can approximate $\rho_{(0,\uparrow)} + \rho_{(0,\downarrow)} \approx 1$. Thus, the current depends only on the difference \mbox{$\Delta \rho \equiv \rho_{(0,\uparrow)}-\rho_{(0,\downarrow)}$}. Lastly, due to the detailed balance of the transition rates~\eqref{Eq:Detailed_Balance}, the current has to vanish for the steady-state population \mbox{$\Delta\rho_\mathrm{th}(t) = -\tanh(\beta \epsilon_B(t))$}. This implies that the current must take the form
\begin{equation}
	J(t) \approx \frac{\Gamma_\mathrm{eff}(t)}{2}(\Delta \rho -\Delta \rho_\mathrm{th}(t)),
\label{Eq:Prob_Current_2}
\end{equation}
where $\Gamma_\mathrm{eff}(t)$ is an instantaneous effective spin-flip rate, which we explicitly derive below.

\noindent The population difference $\Delta \rho$ is then found by solving the differential equation \mbox{$\Delta \dot \rho = -2J(t)$},
\begin{equation}
\begin{split}
	\Delta& \rho(t) = \exp \left[ -\int_{-\infty}^t \mathrm{d}t'\, \Gamma_\mathrm{eff}(t')\right] \bigg( \Delta \rho(-\infty)   \\
	&+ \int_{-\infty}^{t} \mathrm{d}t' \Delta\rho_\mathrm{th}(t') \Gamma_\mathrm{eff}(t') \exp \left[ + \int_{-\infty}^{t'} \mathrm{d}t'' \, \Gamma_\mathrm{eff}(t'') \right] \bigg).
\end{split}
\label{Eq:Diff_Eq_Delta_rho1}
\end{equation}
This result essentially differs from that of Ref.~[\,\onlinecite{Leuenberger2000a}] by the second line, which allows for the electronic spins to equilibrate beyond equal population, i.e. $\Delta \rho(t=\infty) <0$. This is a consequence of our rate equation~\eqref{Eq:Rate_Eq_2} admitting a thermal steady state, while that of Ref.~[\,\onlinecite{Leuenberger2000a}] incorrectly tends to equal populations of the two electronic sectors, even for $B_z \neq 0$.

\subsection{Numerics of dipolar interaction effects}
\label{Sec:Dipolar_Interaction}

Here we briefly describe the algorithm to evaluate the numerics of dipolar interacting spins. We simulate $N_s = 1000$ continuously distributed spins in a cube with dimension $L = N_s^{\frac{1}{3}} = 10$. The mutual interaction of spins $i$ and $j$ is $J_{ij} = s_is_j\sigma_{ij}/r_{ij}^3$, where $s_i,s_j = \pm1$ denote the states of the spins. For simplicity we neglect the angular dependence of the dipolar interaction, instead randomly choosing the sign $\sigma_{ij} = \pm1$ of the interaction for each pair of spins. We include (quasi-) periodic boundary conditions in the sense that we define $r_{ij} = \left[\sum_{\alpha=x,y,z}\left(r_{ij}^{(\alpha)} mod (L/2)\right)^2\right]^{1/2}$ via the Cartesian components $r^(\alpha)_{ij}$ of the vector $\vec{r_{ij}}$. The typical dipolar
interaction is defined as $J_{typ} = 2\pi^2/3$
.

\noindent The local field of a spin $s_i$ at time $t$ consists of the external field – swept
at rate $v$ – and the sum of the dipolar fields, $B_i(t) = vt + \sum_j J_{ij}s_j$. The spins attempt activated tunneling with a rate $\kappa_{exc}$, the attempt of spin $i$ being successful with a Lorentzian probability. The system is initialized at time $t_0 = -10J_{typ}$ with all spins in the up state $(s_j = 1 \forall j)$. The spin-flip probability is then simulated as follows:
\begin{enumerate}
    \item Update the fields of the spins at time $t_0, B_i(t_0) = vt_0 + \sum_j J_{ij}s_j$.
    \item Randomly draw a time step $dt$ from an exponential distribution with
inverse scale parameter $\kappa_{exc}N_s$. Update the time $t_0 \rightarrow t_0 + dt$.
    \item Randomly draw spin $i$, which attempts activated tunneling. The attempt
is successful $(s_i \rightarrow -s_i)$ with probability $\Delta^2/\left(2\left[\Delta^2+\left(2B_i(t_0)\right)^2\right]\right)$.
    \item Repeat steps 1--3 until the final time is exceeded, $t > t_f$.
    \item The spin-flip probability is calculated as the probability to be in the
down-state, $P_{flip} = \frac{1}{N_s}\sum_i\delta_{s_i},-1$.
\end{enumerate}

\begin{figure}[H]
    \centering
	\subfloat[]{
        \includegraphics[width=0.45\linewidth,valign=t]{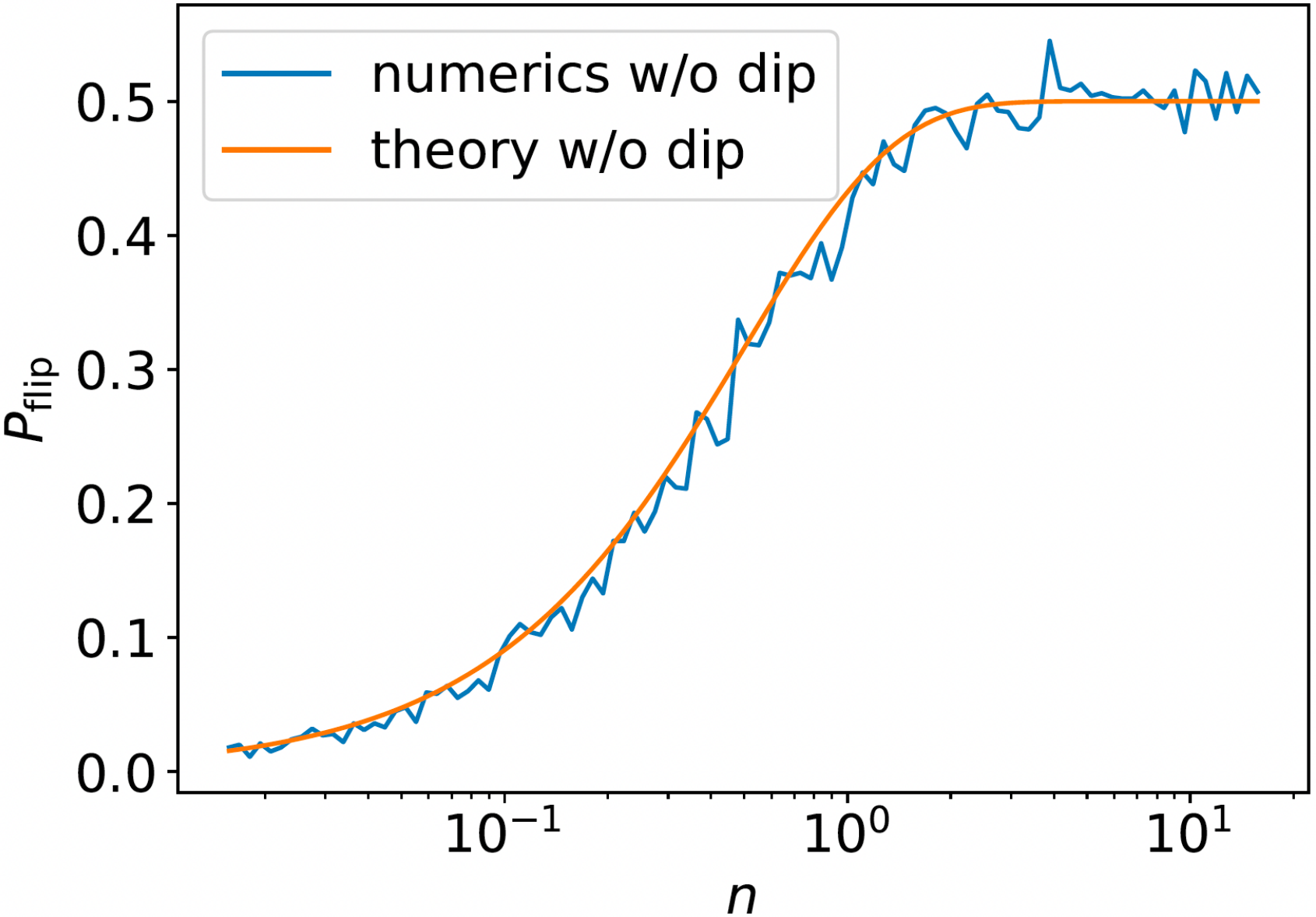}
        \label{Fig:Spin_Flip_wo_Dip}
    }
	\subfloat[]{
        \includegraphics[width=0.5\linewidth,valign=t]{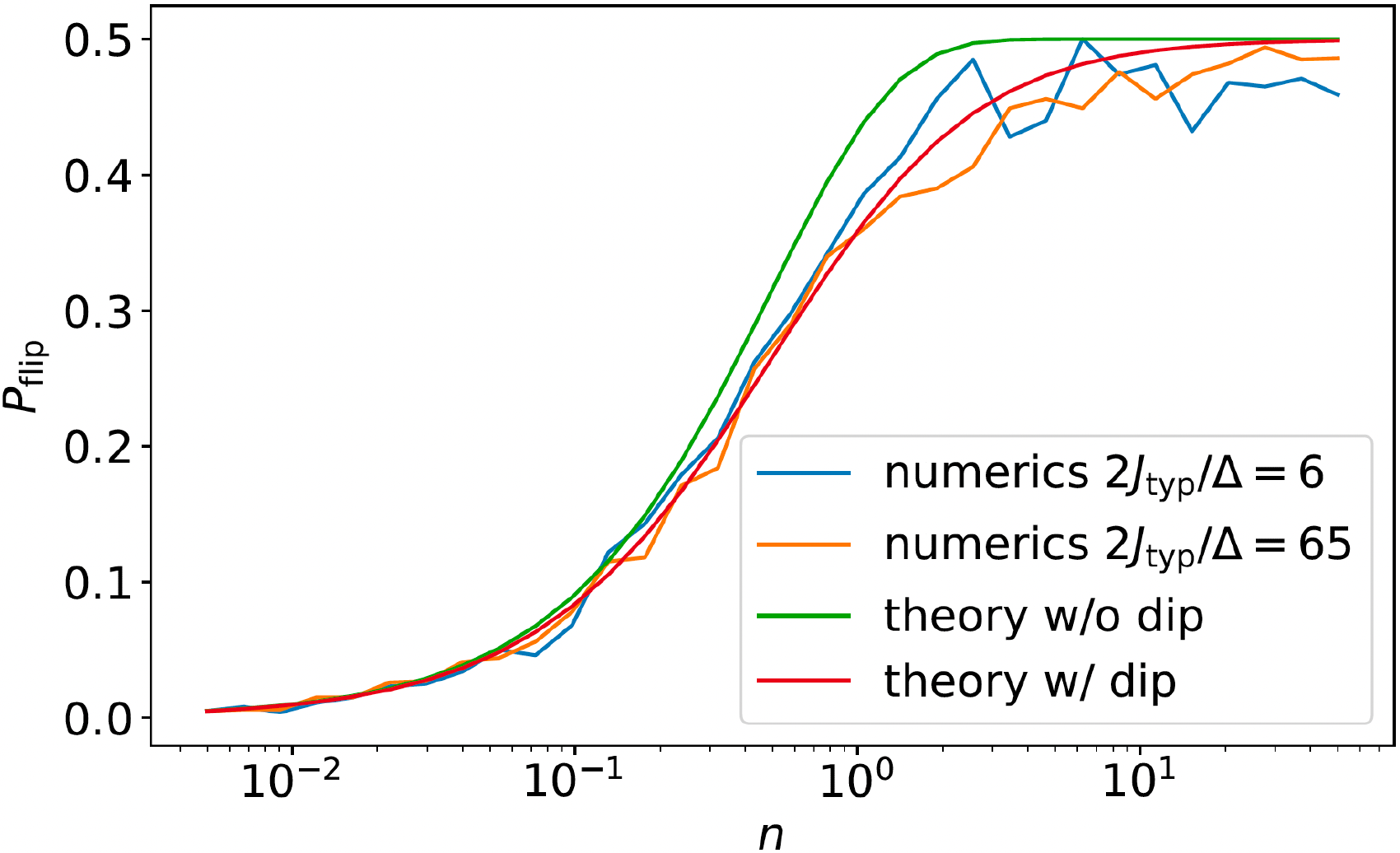}
        \label{Fig:Spin_Flip_with_Dip}
    }
    \caption{Numerically calculated spin-flip probability. \protect\subref{Fig:Spin_Flip_wo_Dip} $P_{flip}$ (blue) without dipolar
interactions plotted against the average number of (successful) spin-flips $n = (1/2)\pi\Delta\kappa_{exc}/(2v)$. The orange curve is the single-ion spin-flip probability
as derived in Eq. \eqref{Eq:P_0Gs}. \protect\subref{Fig:Spin_Flip_with_Dip} With dipolar interaction. The blue and orange curves are calculated for the values $N = 2J_{typ}/\Delta \in \{6,65\}$, showing that the probability does not depend on $N$ (for $N\ll 1$). The green curve is the single-ion spin-flip probability. The red curve shows the phenomenological function (3.24) with $n_{pas} = 2.8$.}
    \label{Fig:Spin_Flip}
\end{figure}

In Fig. \ref{Fig:Spin_Flip_wo_Dip} we show the spin-flip probability as calculated without dipolar interactions, setting $J_{ij} = 0$. As expected, the simple model recovers the analytical result of the single-ion spin-flip probability \eqref{Eq:P_0Gs}. The spin-flip probability with dipolar interactions (3.24) is shown in Fig. \ref{Fig:Spin_Flip_with_Dip}.

\section{Relaxation of nuclear spins due to  coupling to phonons}
\label{Sec:n_phononcoupling}

Here we estimate the rate of phonon emission of hyperfine excitations on Ho ions. 
The decay rate of the first excited CF singlet due to phonon emission was calculated in Ref.~[\onlinecite{Bertaina2006}] $\tau_s^{-1} \approx (1-3) \cdot 10^5/s$. The emission from a nuclear spin excitation involves an additional virtual hyperfine-mediated excitation that flips the nuclear spin and transfers the electronic ground state to the  CF singlet. Using the scaling of the acoustic phonon density of states and of the phonon matrix elements with $\omega^2$ and $\sqrt{\omega}$, respectively, and accounting  for the additional HF transition matrix element $\approx A_{\rm HF}/\Delta$, one obtains  
a nucelar spin-phonon relaxation rate 
$\tau_n^{-1} \approx \left(\frac{E_{\rm HF}}{\Delta}\right)^3 \left(\frac{A_{\rm HF}}{\Delta}\right)^2 \tau_s^{-1}=O(10^{-3}s^{-1})$.
This estimate rationalizes the long persistence of elevated nuclear spin temperatures. At the same time it rules out that phonons play a significant role in the thermally assisted tunneling of the electron spins at level crossings.

It  was shown in Ref.~[\onlinecite{Bertaina2006}] that  a transverse field $B_x$ increases the relaxation rate $\tau_s^{-1}$, which accordingly translates to a faster nuclear spin relaxation.

\section{Dynamics of nuclear spin excitations between Ho ions}
\label{Sec:Ho_Dynamics}

The dipolar interactions and transverse hyperfine interactions can be combined to yield an effective hopping of hyperfine excitations from Ho to Ho ion. For typical Ho ion distances this third-order matrix element scales as $t_{\rm hop}\approx x \cdot 1.8 {\rm MHz}= 5.4{\rm kHz}$ for $x=0.003$.

This hopping competes with two ingredients:
(i) the inhomogeneous linewidth of the 
HF transitions, and (ii) the difference in transition energies for different hyperfine transitions. 

Indeed, the energy of a nuclear spin state with projection $I_z$ on the Ho's electronic spin is well described by 
\begin{equation}
    E(I_z) = const. + A I_z + B I_z^2. 
\end{equation}
The linear term is predominantly due to the coupling to the magnetic moment of the electronic spin and the Zeeman field, while the quadratic contributions are due to quadrupolar interactions as well as second order virtual processes due to the leading $I\cdot J$ hyperfine interaction. In general the coefficient $A$ is different for nuclear spin states of up or down pointing electronic spins. Moreover it varies from ion to ion due to internal field disorder, which contributes on the order of $20{\rm kHz}$ to variations $\delta A$ at the concentration $x=0.003$. A larger inhomogeneous broadening is expected from (e.g. strain-induced) variations of the magnetic moment of the electronic spin of Ho ions. We estimate its relative variations to be on the order of $2\cdot 10^{-5}$, extrapolating from strain-induced shifts of crystal field levels in the same host material doped with Tb~\cite{Beckert:2024pi}. This  translates to inhomogeneous broadening of the hyperfine transitions of $\delta A \approx 80 {\rm kHz}$.

$B$ is of the order of $15$MHz. The hyperfine transition energies of adjacent excitation levels, $I_z\leftrightarrow I_{z+1}$ and $I_{z+1}\leftrightarrow I_{z+2}$, thus differ by about $2B= 30 MHz$, which is far bigger than the matrix element for the dipolar exchange of a hyperfine excitation. 

\subsection{Diffusion of nuclear spin excitations}

Given the comparatively large value of $B$ the only nuclear spin processes that are potentially resonant, but involve only a single dipolar coupling, are those in which the ions swap their electro-nuclear state. This requires that they are in the same electronic spin state (unless $B_z\approx 0$), and that the nuclear spin projections on the respective electron spins differ by 1 unit. 
In such processes ion undergoes a transition $I_z\leftarrow I_{z+1}$, while the other one undergoes the reverse transition $I_{z+1}\leftarrow I_{z}$. While the quadratic contribution $O(B)$ drops out, there is still a disorder of order $\delta A$  to overcome. In the absence of a bath the only algebraically decaying dipolar interactions would still lead to dynamics, but it would be strongly suppressed by powers of $t_{\rm hop}/\delta A$.  

However, the nuclear spins of the fluorine ions of the LiYF$_4$ host from a nuclear spin bath with a width of order $O(100-150)$kHz. Dipolar coupling of one of the involved Ho ions to the fluorine bath then allows to bridge the energy mismatch of order $\delta A$. A rough estimate of the Fermi Golden rule rate of such assisted hopping processes yields a hopping rate of $O(100)$Hz.

Note that these hopping processes facilitate the diffusion from ion to ion of existing hyperfine excitations in the material, and thus enables thermally assisted tunneling of those ions that undergo an anticrossing. However, the diffusion preserves the total number of excitations with given $I_z$.

\subsection{Thermalization of nuclear spin excitations on Ho ions}

To equilibrate the population of hyperfine excitations requires nuclear spin processes that change the overall hyperfine population. Interestingly, there exist such processes, involving 3 ro 4 ions, that do not rely on the very weak coupling to phonons to achieve such a relaxation.

To change the hyperfine state of a set of $n$ Ho ions, $n$ transverse hyperfine interactions and $n-1$ dipolar interactions among the electronic spins must combine in a higher order process. Each dipolar interaction can change $I_z$ of the involved Ho ions by $\Delta I_z^i = \pm 1$ or $0$, whereby the total number of hyperfine excitation units remains constant ($\sum_{i=1}^n \Delta I_z^i = 0$), so as to cancel the leading hyperfine energy change $\propto A$.
The process must further be such that the quadratic contributions $O(B)$ to the hyperfine energy cancel.
Consider $n=4$ ions, and let ions 1 and 2 increment their $I_z$ by 1, while ions 3 and 4 decrement it by 1.
The constraint $I_z^1+I_z^2-I_z^3-I_z^4 = -2$ then assures the sought cancellation, and only energy mismatches due to inhomogeneities $\delta A_i$ remain, which are small enough to be bridged by the fluorine bath. 

There are also some resonant processes that involve only 3 ions, but they are too constrained to induce full equilibration. The number of distinct transitions of $n=4$ ions satisfying the above constraint is instead sufficient to ensure full equilibration with no involvement of the external bath. The rate of such fluorine assisted collective flip-flop processes is expected to be of order $O(1-10s)$, which is still significantly faster than relaxation involving relaxation through the phonon bath.  

These phonon-free processes lead to ergodicity within the nuclear spin subsystem (in general only among the Ho ions with the same electron spin polarization though), establishing a nuclear spin temperature $T_n$.
Since the full Ho nuclear spin population equilibrates to a common $T_n$ during the fast and fully mixing spin dynamics that set in at the level crossings $B_z=0$Gs and$B_z= 230$Gs, it is reasonable to expect that the two a priori different nuclear spin temperatures pertaining to the up and the down electron spin sector do not differ by much across the entire hysteresis curve.


\end{document}